\documentclass[12pt,draftclsnofoot, onecolumn]{IEEEtran}
\usepackage{array}
\usepackage{graphicx}
\usepackage{epsfig}
\usepackage{cite}
\usepackage{epstopdf}
\usepackage{amsfonts,amsmath,amsthm,amssymb,balance}
\usepackage{multirow,balance}
\usepackage{bm}
\usepackage[usenames,dvipsnames]{color}
\usepackage{colortbl}
\usepackage{float}
\usepackage{subfigure}
\usepackage{dcolumn}
\usepackage{cuted}
\usepackage{changepage}
\usepackage[ruled]{algorithm2e}                 

\definecolor{mygray}{gray}{.9}

\graphicspath{{figures/}}
\newcolumntype{C}[1]{>{\PreserveBackslash\centering}p{#1}}
\newcolumntype{R}[1]{>{\PreserveBackslash\raggedleft}p{#1}}
\newcolumntype{L}[1]{>{\PreserveBackslash\raggedright}p{#1}}

\newtheorem{definition}{Definition}

\usepackage[flushleft]{threeparttable}

\begin{document}

\title{Uniquely Decodable Multi-Amplitude Sequence for Grant-Free Multiple-Access Adder Channels}

\vspace{50pt}

\author{Qi-Yue~Yu, ~\IEEEmembership{Senior~Member,~IEEE},
        and~Ke-Xun~Song

}


\markboth{IEEE Transactions on Information Theory, ~Vol.~XX, No.~X, X~2022}%
{Shell \MakeLowercase{\textit{et al.}}: Bare Demo of IEEEtran.cls for IEEE Journals}

\maketitle
\begin{abstract}
Grant-free multiple-access (GFMA) is a valuable research topic, since it can support multiuser transmission with low latency. 
This paper constructs novel uniquely-decodable multi-amplitude sequence (UDAS) sets for GFMA systems, 
which can provide high spectrum efficiency (SE) with low-complexity active user detection (AUD) algorithm. 
First of all, we propose an UDAS-based multi-dimensional bit interleaving coded modulation (MD-BICM) transmitter; then introduce the definition of UDAS and construct two kinds of UDAS sets based on cyclic and quasi-cyclic matrix modes. Besides, we present a statistic of UDAS feature based AUD algorithm (SoF-AUD), and a joint multiuser detection and improved message passing algorithm for the proposed system. 
Finally, the active user error rate (AUER) and Shannon limits of the proposed system are deduced in details. Simulation results show that our proposed system can simultaneously support four users without additional redundancy, and the AUER can reach an extremely low value $10^{-5}$ when $E_b/N_0$ is $0$ dB and the length of transmit block is larger than a given value, i.e., 784, verifying the validity and flexibility of the proposed UDAS sets.
\end{abstract}


%
\IEEEpeerreviewmaketitle

 \newpage
 \setcounter{page}{1}

\section{Introduction}
%
%
%
%

\IEEEPARstart{N}{ext} generation multiple-access is expected to support massive users in the limited resources, and many works have been done on this topic \cite{Dai2015,Ding2017,Dai2016}.
In general, the multiple-access technique can be categorized into uncoordinated multiple-access and coordinated multiple-access \cite{RandomAccess1_Wu_2020}. 
The uncoordinated multiple-access is absent coordination and generally viewed as unsourced multiple-access,
in which each user shares the same transmission protocol without allocated signature before transmission.
Conversely, the coordinated scenario is to administer the users by a central processor, i.e., base station (BS), and each user is assigned some unique signature that can be recognized by the receiver for detection. 

The grant-free multiple-access (GFMA) can be viewed as a special case of the coordinated multiple-access, where each user may access the BS randomly.
The BS needs to detect both the number of active users and their corresponding data sequences.
The major difference between a grant-free coordinated multiple-access and an uncoordinated multiple-access is relayed on the dedicated signature \cite{RandomAccess1_Wu_2020}.
The grant-free scenario generally allocates a pilot sequence (or signature) to each user; thus, the receiver can separate and identify the users with the help of pilot sequences.
In contrast, the active users of the uncoordinated multiple-access case randomly select pilot sequences without any coordination, and sometimes may lead to pilot collisions.

Therefore, it is important to design multiuser signatures for GFMA systems.
The well-known pseudo random sequences are composed of binary bits $\{0, 1\}$, e.g., m-sequence, golden sequences \cite{Golden_2005}, Reed-Muller codes \cite{RM_2018}, Walsh sequences, and etc.
With the development of spreading sequences, many papers discuss uniquely-decodable (UD) ternary code sets $\{-1, 0, +1\}$ for the overloaded synchronous code division multiple-access (CDMA) systems \cite{UD_CDMA1_2012, UD_CDMA2_2014, UD_CDMA3_2019, UD_CDMA4_2012, UD_CDMA5_2016, UD_CDMA6_2018}, which can support larger number of users than the classical orthogonal spreading codes. 
Besides the binary and ternary code sequences, Zad-Off Chu (ZC) sequences are also popular pilot sequences, especially for channel estimation. 
It is found that the amplitude of ZC sequences is generally a constant, and only phases are varied \cite{Popovic1992, Hua2013, Tsai2013}.
In addition, some papers are interested in designing frameworks to avoid collisions for GFMA systems, 
e.g., paper \cite{GF_2017} treats collisions as interference and builds the statistical model with the aid of Poisson point processes.
Moreover, multiuser codebook design is also taken into consideration for GFMA.
In \cite{GF_SCMA1_2014}, it presents multi-dimensional (MD) codebooks of sparse code multiple access (SCMA) in a grant-free multiple-access channel (MAC), where the data stream of each user is directly mapped to a codeword of the proposed MD codebook.
Due to the sparsity of MD codebooks, the proposed scheme can maintain overloaded information and enable massive connectivity.
When these classical pilot sequences (or MD codebooks) are used as multiuser signatures for a GFMA system, there exist some challenges and/or drawbacks.
\begin{itemize}
	\item
	The spectrum efficiency (SE) of spreading sequences based multiple-access systems is generally low. 
	For example, when a binary spreading sequence (e.g. Walsh sequence) is utilized, the SE of each user is equal to the reciprocal of the spreading factor (SF), i.e., $\frac{1}{SF}$, and the sum-rate of the multiuser system is equal to or smaller than one.
	When UD ternary codes are utilized, the sum-rate can be larger than one, because of the overloaded information; however, the SE of each user is still equal to $\frac{1}{SF}$. 
	\item
	The designed multiuser codebooks cannot be flexibly extended to a general multiple-access case. Most of the recent multiuser codebooks are designed based on multi-dimensional constellations (or lattice, and etc.), which have strict constraints on transmit signals' amplitudes and phases \cite{GF_SCMA_Codebook1_2017,GF_SCMA_Codebook2_2018,GF_SCMA_Codebook3_2019,GF_SCMA_Codebook4_2017,GF_SCMA_Codebook5_2018}. Thus, the extension of the designed multiuser codebooks is generally insignificant, especially for a massive random access system.
	\item
	Most of the recent active user detection (AUD) algorithms of GFMA systems are with relatively high complexities, and few of them focus on the theoretical analyses of the AUD processing.
	For example, the primary AUD methods are compressed sensing techniques \cite{GF_CS1_2016, GF_CS2_2017, RandomAccess3_Wu_2018}, successive joint decoding (SJD) \cite{GF_2017}, successive interference cancellation (SIC), blind detections \cite{GF_SCMA1_2014}, and etc. 
\end{itemize}

Regarding as the aforementioned challenges, it is interesting to design special sequences for GFMA systems.
Until now, most of the classical pilot sequences are designed based on binary (or ternary, or phase) sets, few concerns on the multi-amplitude information.
In \cite{UDM_Yu_2019}, we have proposed the concept of uniquely-decodable mapping (UDM).
It is declared that, if each user exploits 2ASK (amplitude shift keying) and the amplitudes of $J$ users are respectively $\{1, 2^1, \ldots, 2^{J-1}\}$, then the $J$ users can be uniquely separated without ambiguity at the receiver. For example, if there are two users and the transmit signals of the two users are respectively $\{-1, +1\}$ and $\{-2, +2\}$, the superimposed signal set is $\{-3, -1, +1, +3\}$ that is a one-to-one mapping between the two users' transmit signals and the superimposed signals.

Based on the conception of UDM, this paper constructs a novel uniquely-decodable multi-amplitude sequence (UDAS) for GFMA systems. The advantages of the proposed UDAS include three aspects. First of all, it presents a high SE without additional redundancy, likewise, some designed non-orthogonal multiple-access (NOMA) codebooks.
Following, it can be easily applied to various multiple-access scenarios, with favoured flexible.
Finally, the AUD algorithm can be easily realized.
The major contributions of this paper are four-fold:
\begin{enumerate}
\item
This paper firstly proposes an UDAS-based multi-dimensional bit interleaving coded modulation (MD-BICM) transmitter, which is a combination of two channel encoders, one interleaver and multi-dimensional modulation.
\item
We define the conception of UDAS, and construct two kinds of UDAS sets based on cyclic and quasi-cyclic matrix modes. Some important features of the cyclic/quasi-cyclic UDAS sets are deduced in details.
\item
We propose a statistic of UDAS feature based AUD algorithm (SoF-AUD), and a joint multiuser detection (MUD) and improved message passing iterative decoding algorithm for the proposed system.
\item
Finally, we deduce the theoretical active user error rate (AUER) of the proposed system;
then, analyze and calculate the Shannon limits of the multiple-access adder channels in details.
\end{enumerate}

The remainder of this paper is organized as follows. 
In Section II, the UDAS-based MD-BICM system is described.
We present some definitions of UDAS, and construct two kinds of UDAS sets with detailed features in Section III.
Section IV proposes detection algorithms for the proposed system.
The theoretical analyses are presented in Section V.
Simulation results are discussed in Section VI, followed by concluding remarks drawn in Section VII.

\section{System model}\label{section2}
In this paper, $a$, $\mathbf{a}$ and $\mathbf{A}$ stand for a variable, a vector and a matrix, respectively. Denote ${\bf A}^{\rm T}$ by the transpose of a matrix ${\bf{A}}$.
Let $\mathbb {B}$, $\mathbb {Z}$ and $\mathbb {C}$ be the binary, integer and complex fields, respectively.
${\rm Re}[a]$ and ${\rm Im}[a]$ are respectively the real part and image part of the complex number $a$.
$1i$ stands for an image number.
${\rm E}[.]$ is the function of expectation, and $\lceil . \rceil$ denotes the ceiling operator.
$\left\| {\bf a} \right\|_2$ denote the 2-norm of a vector ${\bf a}$.

Suppose there are $J$ users simultaneously access the BS. 
Obviously, the value of $J$ should be estimated for a GFMA system.
Each user is equipped with a single antenna, and the BS holds one antenna.
The system model is shown in Fig. 1.

  \begin{figure*}[t]
    \centering
    \includegraphics[width=0.95\linewidth]{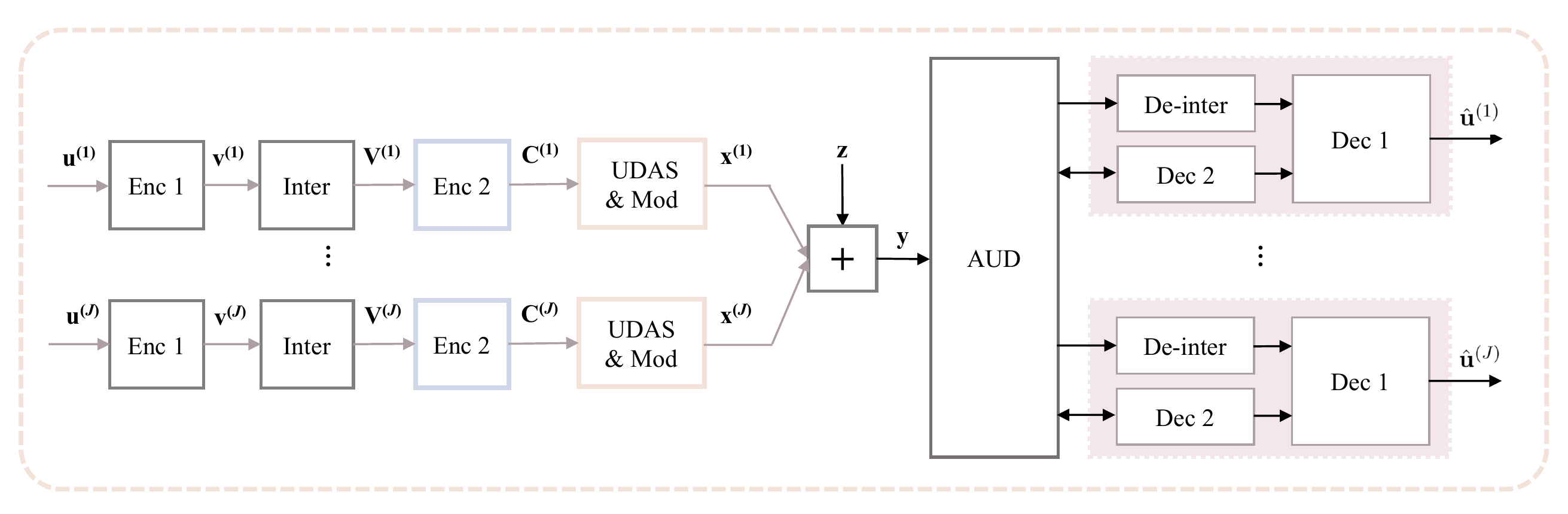}
    \caption{A diagram of the proposed UDAS-based MD-BICM system, where ``Enc'', ``Dec'', ``Inter'', ``De-inter'', and ``Mod'' respectively stand for ``encoder'', ``decoder'', ``interleaver'', ``de-interleaver'' and ``modulation''.}
    \label{Fig1}
    \vspace{-20pt}
    \end{figure*}

The transmit bit information of the $j$th user is defined by ${\bf u}^{(j)} \in {\mathbb B}^{1 \times K}$, 
where $1 \le j \le J$ and $K$ is the number of bits of the $j$th user.
${\bf u}^{(j)}$ is then passed to the first channel encoder whose generator is a $K \times N_1$ matrix defined by ${\bf G}_1 = [g_{k,n_1}]_{1 \le k \le K, 1 \le n_1 \le N_1}$, 
and obtained an encoded codeword ${\bf v}^{(j)} = {\bf u}^{(j)} \cdot {\bf G}_1$, 
where ${\bf v}^{(j)} = (v_1^{(j)}, v_2^{(j)},\ldots,v_{n_1}^{(j)},\ldots, v_{N_1}^{(j)})$. 
We can write ${\bf v}^{(j)}$ into a matrix ${\bf V}^{(j)}$ column-by-column, given as
\begin{equation*}
{\bf V}^{(j)} = \left[
	\begin{matrix}
		v_1^{(j)} & v_{M+1}^{(j)} & \ldots & v_{(N_c-1)M+1}^{(j)}\\
		v_2^{(j)} & v_{M+2}^{(j)} & \ldots & v_{(N_c-1)M+2}^{(j)}\\
		\vdots & \vdots & \ddots & \vdots\\
		v_M^{(j)} & v_{2M}^{(j)} & \ldots & v_{N_c M}^{(j)}\\
	\end{matrix}
	\right],
\end{equation*}
which is an $M \times N_c$ matrix with the condition $N_1 \le M \cdot N_c$, and $v_{n_1}^{(j)} = 0$ for all $N_1 < n_1 \le M \cdot N_c$.
Actually, ${\bf V}^{(j)}$ is utilized to show the interleaving process, 
since the encoded codeword ${\bf v}^{(j)}$ will be further processed row-by-row, and the number of columns in ${\bf V}^{(j)}$ is directly related to the length of the selected UDAS.

Then, each row of ${\bf V}^{(j)}$ selects $L-1$ bits, 
and passes the selected $L-1$ bits to the second encoder ${\bf G}_2$.
In fact, ${\bf G}_2$ is designed for AUD, and it can be as simple as possible, 
e.g., single parity-check code (SPC).
For the $m$th row of ${\bf V}^{(j)}$ for $1 \le m \le M$, 
let the parity-check bit generated by ${\bf G}_2$ be $v_{m,pc}^{(j)}$, 
which is located at the end of the $m$th row of ${\bf V}^{(j)}$. 
Thereafter, we can achieve an $M \times N$ matrix ${\bf C}^{(j)}$, as
\begin{equation*}
{\bf C}^{(j)} = \left[
	\begin{matrix}
		{\bf c}_1^{(j)}\\
		{\bf c}_2^{(j)}\\
		\ldots\\
		{\bf c}_M^{(j)}\\
	\end{matrix}
	\right] =	
	\left[
	\begin{matrix}
		v_1^{(j)} & v_{M+1}^{(j)} & \ldots & v_{(N_c-1)M+1}^{(j)} & \textcolor{blue}{v_{1,pc}^{(j)}}\\
		v_2^{(j)} & v_{M+2}^{(j)} & \ldots & v_{(N_c-1)M+2}^{(j)} & \textcolor{blue}{v_{2,pc}^{(j)}}\\
		\vdots & \vdots & \ddots & \vdots & \vdots\\
		v_M^{(j)} & v_{2M}^{(j)} & \ldots & v_{N_c M}^{(j)} & \textcolor{blue}{v_{M,pc}^{(j)}}\\
	\end{matrix}
	\right],
\end{equation*}
where ${\bf c}_m^{(j)} = \left(v_m^{(j)}, v_{M+m}^{(j)}, \ldots, v_{(N_c-1)M+m}^{(j)}, 
\textcolor{blue}{v_{m,pc}^{(j)}}\right)$ is a ${1\times N}$ vector, with $N = N_c+1$.

Consequently, ${\bf C}^{(j)}$ is modulated row-by-row.
Take the $m$th row ${\bf c}_{m}^{(j)}$ as example to explain the modulation mapping. 
Assume the multi-dimensional (MD) modulation index is $\mathcal M$ and the length of an UDAS is $L$,
under the assumption $N = L \cdot \log_2 \mathcal M$.
For further discuss convenience, we rewrite ${\bf c}_m^{(j)}$ as
\begin{small}
\begin{equation*}
	\begin{aligned}
{\bf c}_m^{(j)} &= \left( \{c_{m,1,1}^{(j)}, c_{m,1,2}^{(j)}, \ldots, c_{m,1,{\log_2 \mathcal M}}^{(j)}\}, \ldots, 
	\{c_{m,l,1}^{(j)}, c_{m,l,2}^{(j)}, \ldots, c_{m,l,{\log_2 \mathcal M}}^{(j)}\}, \ldots,
	\{c_{m,L,1}^{(j)}, c_{m,L,2}^{(j)}, \ldots, c_{m,L,{\log_2 \mathcal M}}^{(j)}\} \right)\\
				&= \left( {\bf c}_{m,1}^{(j)},\ldots, {\bf c}_{m,l}^{(j)}, \ldots, {\bf c}_{m,L}^{(j)} \right)
	\end{aligned}
\end{equation*} 
\end{small}
where $c_{m,l,b}^{(j)} \in \mathbb{B}$ for $1 \le l \le L$ and $1 \le b \le \log_2{\mathcal M}$,
and ${\bf c}_{m,l}^{(j)} = \{c_{m,l,1}^{(j)}, c_{m,l,2}^{(j)}, \ldots, c_{m,l,{\log_2 \mathcal M}}^{(j)}\}$.
Then, every $\log_2 {\mathcal M}$ bits of ${\bf c}_{m}^{(j)}$ are modulated to one symbol,
i.e., ${\bf c}_{m,l}^{(j)} \to x_{m,l}^{(j)}$.
We can achieve a modulated $M \times L$ matrix ${\bf X}^{(j)}$ as
\begin{equation*}
{\bf X}^{(j)} = \left[
	\begin{matrix}
		{\bf x}_1^{(j)}\\
		{\bf x}_2^{(j)}\\
		\ldots\\
		{\bf x}_M^{(j)}\\
	\end{matrix}
	\right] =	
	\left[
	\begin{matrix}
		x_{1,1}^{(j)} & x_{1,2}^{(j)} & \ldots & x_{1,L}^{(j)}\\
		x_{2,1}^{(j)} & x_{2,2}^{(j)} & \ldots & x_{2,L}^{(j)}\\
		\vdots & \vdots & \ddots & \vdots \\
		x_{M,1}^{(j)} & x_{M,2}^{(j)} & \ldots & x_{M,L}^{(j)}\\
	\end{matrix}
	\right],
\end{equation*}
where $x_{m,l}^{(j)}$ is an ${\mathcal M_1}$-dimensional modulated symbol, where $\mathcal M = 2\mathcal M_1$, expressed as
\begin{equation*}
	\begin{aligned}
		(s_{m,l}^{(j)}, 0, 0, \ldots, 0),
		(0, s_{m,l}^{(j)}, 0, \ldots, 0),
		&\ldots,
		(0, 0, 0, \ldots, s_{m,l}^{(j)}),
	\end{aligned}
\end{equation*}
in which only one position of $x_{m,l}^{(j)}$ has value $s_{m,l}^{(j)}$ and the other ${\mathcal M_1}-1$ positions are all zeros, for $1 \le m \le M$ and $1 \le l \le L$.
Note that ${\mathcal M_1}$-dimensional modulation can increase both the flexibility and the minimum Euclidean distance, thus is appealing for GFMA.
If the $i$th location of $x_{m,l}^{(j)}$ is non-zero, we set the location index $\rho_{m,l,i}^{(j)} = 1$,
and other location indexes $\rho_{m,l,i'}^{(j)} = 0$ for $i' \neq i$ and $1 \le i' \le \mathcal M_1$.
The non-zero location is determined by the first $\log_2 \mathcal{M}_1$ bits of ${\bf c}_{m,l}^{(j)}$.

Suppose the $j$th user utilizes ${\bf e}_j =(e_{j,1}, e_{j,2},\ldots, e_{j,L}) $ as its UDAS, 
then $s_{m,l}^{(j)}$ is calculated as
\begin{equation} \label{e_s_mlj}
{s}_{m,l}^{(j)} = \left( 2 c_{m,l,\log_2{\mathcal M}}^{(j)}-1 \right) \cdot {e}_{j,l},
\end{equation}
which is determined by the last bit ${c}_{m, l \cdot \log_2{\mathcal M} }^{(j)}$ of ${\bf c}_{m,l}^{(j)}$ 
and the $l$th symbol of ${\bf e}_j$. 

Look back to the second encoder ${\bf G}_2$. 
Since ${\bf G}_2$ is used to assist AUD, the input $(L-1)$ bits of ${\bf G}_2$ are set to be the last bit of ${\bf c}_{m,l}^{(j)}$ for $1 \le l \le L$, then
\begin{equation*}
	\begin{aligned}
	v_{m,pc}^{(j)} &= c_{m,1,\log_2{\mathcal M}}^{(j)} \oplus c_{m,2,\log_2{\mathcal M}}^{(j)}
	               \oplus \ldots \oplus c_{m,L,\log_2{\mathcal M}}^{(j)}\\
	               &= v_{m + M \cdot (\log_2 {\mathcal M} - 1)}^{(j)} \oplus v_{m + M \cdot (2\log_2 {\mathcal M} - 1)}^{(j)} \oplus \ldots \oplus v_{m + M \cdot ((L-1) \log_2 {\mathcal M} - 1)}^{(j)}.
	\end{aligned}
\end{equation*}

For better understand, an example is presented to show the interleaving and encoding process.

\textbf{Example 1}:
Assume ${\bf v}^{(j)} = (0, 1, 1, 0, 1, 0, 1, 0, 0, 1, 0, 0, 0, 1, 0, 1, 1, 1, 1, 0, 1)$, which is a $1 \times 21$ vector and $N_1 = 21$.
Set $M = 2$, $N_c = 11$, $\mathcal M = 8$ and $\mathcal M_1 = 4$, then ${\bf v}^{(j)}$'s corresponding matrix ${\bf V}^{(j)}$ is given by
\begin{equation*}
{\bf V}^{(j)} = \left[
	\begin{array}{cccccccccccc}
		0 & 1 & {1} & 1 & 0 & {0} & 0 & 0 & {1} & 1 & 1\\
		1 & 0 & {0} & 0 & 1 & {0} & 1 & 1 & {1} & 0 & \textcolor{red}{0}\\
	\end{array}
\right],
\end{equation*}
and $v_{2,11}$ is set to be 0 (shown in red color). Then, SPC is done to each row of ${\bf V}^{(j)}$, given by
$v_{1,p_1}^{(j)} = v_5 \oplus v_{11} \oplus v_{17} = 0$ and $v_{2,p_1}^{(j)} = v_6 \oplus v_{12} \oplus v_{18} = 1$. Thus, it is obtained
\begin{equation*}
{\bf C}^{(j)} = \left[
	\begin{matrix}
		{\bf c}_1^{(j)}\\
		{\bf c}_2^{(j)}\\
	\end{matrix}
\right]
= \left[
	\begin{matrix}
		01\textcolor{blue}{1} & 10\textcolor{blue}{0} & 00\textcolor{blue}{1} & 11\textcolor{red}{0}\\
		10\textcolor{blue}{0} & 01\textcolor{blue}{0} & 11\textcolor{blue}{1} & 00\textcolor{red}{1}\\
	\end{matrix}
\right],
\end{equation*}
which is a $2 \times 12$ matrix, where $N = 12$ and $L=4$. End Example 1.

Then, ${\bf X}^{(j)}$ is sent to the adder multiple-access channel row-by-row.
At the receiver, the received signal of the $m$th row ${\bf y}_m$ is equal to
\begin{equation}
{\bf y}_m = \sum_{j=1}^{J} {\bf x}_m^{(j)} + {\bf z}_m,
\end{equation}
where $1 \le m \le M$, and ${\bf z}_m$ is an additive white Gaussian noise (AWGN) vector.
Each noise component of ${\bf z}_m$ is an independent and identically distributed ({\it i.i.d}) Gaussian random variable with distribution ${\cal {N}}\left( {0,{N_0}/2} \right)$.
Set ${\bf y} = ({\bf y}_1, {\bf y}_2, \ldots, {\bf y}_M)$, 
where ${\bf y}_m = (y_{m,1}, y_{m,2}, \ldots, y_{m,l}, \ldots, y_{m,L})$ includes $L$ symbols.
For the $l$th received symbol $y_{m,l}$ of ${\bf y}_m$, we have
\begin{equation*}
	y_{m,l} = (y_{m,l,1}, y_{m,l,2}, \ldots, y_{m,l,i},\ldots,y_{m,l,\mathcal M_1}),
\end{equation*}
which is an $\mathcal M_1$-dimensional signal, and 
\begin{equation} \label{e4}
y_{m,l,i} = \sum_{j=1}^{J} \rho_{m,l,i}^{(j)} s_{m,l}^{(j)} + z_{m,l,i},
\end{equation}
where $\rho_{m,l,i}^{(j)} = \{0, 1\}$.
Based on the received ${\bf y}$, the AUD modular can detect the number of arrival users and identify users' signatures.
Afterwards, by a joint detection algorithm, all the users' information bits can be recovered.
It is noted that the two encoders (${\bf G}_1$ and ${\bf G}_2$) can be removed, and the system still works, which reflects the great flexibility of the proposed scheme.

In the following discussion, the subscripts ``$j$'', ``$m$'', ``$l$'', ``$b$'' and ``$i$'' still stand for the $j$th user, the $m$th row, the $l$th symbol, the $b$th bit of the symbol, 
and the $i$th location of the $\mathcal M_1$-dimension modulation, 
where $1 \le j \le J, 1 \le m \le M, 1 \le l \le L$, $1 \le b \le \log_2\mathcal M$ and $1 \le i \le \mathcal M_1$.

\section{Definition, construction, and features of UDAS set}
This section presents definition, construction and features of UDAS set in synchronous adder multiple-access channels. A good designed UDAS set can separate multiuser without ambiguous.

\subsection{Definition}
According to \cite{UDM_Yu_2019}, an uniquely-decodable mapping (UDM) element set is defined by ${\Delta} =\{\Delta_{re}, \Delta_{im} \}$, where $\Delta_{re} = \{1, 2,\ldots, 2^p\}$ and $\Delta_{im} = \{1i, 2i,\ldots, 2^p i\}$ with $p \in \mathbb{Z}$.
This UDM element set can maximum simultaneously support $2(p+1)$ users without ambiguous, if each user selects one of the element, i.e., $a$, and its inverse, i.e., $-a$, as the user's modulated symbols \cite{UDM_Yu_2019}.
Denote $|\Delta|$, $|\Delta_{re}|$, and $|\Delta_{im}|$ by the number of elements in sets $\Delta$, $\Delta_{re}$, and $\Delta_{im}$, respectively.
Obviously, it is derived that $|\Delta_{re}| = |\Delta_{im}| = p+1$ and $|\Delta| = 2(p+1)$, which are determined by the parameter $p$.

\begin{definition}
An $L$-length UDAS is consisted of $L$ sequential elements, which are selected from an UDM element set ${\Delta}$ with $p$ as a parameter.
\end{definition}

Let all the $L$-length UDAS belong to a space ${\mathbb E}_L$, and the total number of ${\mathbb E}_L$ space is equal to $|{\mathbb E}_L|=|\Delta|^L = [2(p+1)]^L$.


\begin{definition}
Assume a set $\Psi$ contains $T$ sequences, as
\begin{equation*}
{\Psi} = \{ {\bf e}_1, {\bf e}_2, \ldots,{\bf e}_t , \ldots, {\bf e}_T\},
\end{equation*}
with ${\bf e}_t = (e_{t,1}, e_{t,2}, \ldots, e_{t,l}, \ldots, e_{t,L}) \in {\mathbb E}_L$, where $e_{t,l} \in \Delta$ and $1 \le l \le L$.
There are $T$ users, and each user is assigned a sequence, i.e., ${\bf e}_t$, 
and the transmit vector of the $t$th user is defined by 
${\bf c}^{(t)} = (c_{1}^{(t)}, c_{2}^{(t)}, \ldots, c_{l}^{(t)},\ldots,c_{L}^{(t)})$ 
where ${\bf c}^{(t)} \in {\mathbb B}^{1 \times L}$ for $1 \le t \le T$.
The modulated symbol of the $l$th symbol of the $t$th user is equal to $(2c_l^{(t)} - 1) \cdot e_{t,l}$,
thus the sum-pattern of $T$ users $w_l$ can be given as 
$w_l=\sum_{t=1}^{T} (2c_l^{(t)} - 1) \cdot e_{t,l}$, 
where $w_l \in \Omega_{l}^T$ and  $|\Omega_{l}^T| = 2^T$ for $1 \le l \le L$.
Then, the sum-pattern vector of the entire $L$ symbols of the $T$ users is defined by
${\bf w} = \{w_1, w_2, \ldots, w_l,\ldots, w_L\}$ that belongs to the sum-pattern set $\Omega^T$, 
i.e., ${\bf w} \in \Omega^T$,
where $\Omega^T = \{ \Omega_{1}^T, \Omega_{2}^T, ..., \Omega_{l}^T,\ldots, \Omega_{L}^T\}$.
If the sequences in ${\Psi}$ satisfy the following conditions:
\begin{enumerate}
\item
When $t \neq t'$ and $1 \le t, t' \le T$, we have $e_{t,l} \neq e_{t',l}$ for all $1 \le l \le L$;
\item
All the vectors in $\Omega^{T}$ are different from each other;
\item
It is a one-to-one mapping between the sum-pattern vector and the transmit vectors of $T$ users, 
i.e., ${\bf w} \leftrightarrow \{ {\bf c}^{(1)}, {\bf c}^{(2)}, \ldots,{\bf c}^{(t)},\ldots,{\bf c}^{(T)} \}$; and
\item
The power constraint satisfies $\lim_{L \to \infty} {\mathbb P} \left( | {P}_{t} - P_{avg}| < \varepsilon \right) = 0$, where $P_{t} = \frac{1}{L} \sum_{l=1}^{L} e_{t,l}^2$ stands for the average power of the $t$th UDAS, $P_{avg} = \frac{1}{T} \sum_{t=1}^{T} P_t$ is the average power of the entire UDAS set, and $\varepsilon$ is a positive small number;
\end{enumerate}
then, the set ${\Psi}$ is defined by a $T$-size UDAS set.
\end{definition}

The former three conditions are used to ensure the superimposed signals of $T$ users can be uniquely separated.
The last condition anticipates the average power of each user keeps as a constant.
For a multiple-access network, we hope the average transmit power of each user is almost the same, 
so that to keep fairness among users.
Besides the aforementioned conditions, there are some supplementary conditions, for example,
$\frac{  \left\| {\bf e}_t \right\|_\infty }{\sqrt{P_{avg}}}  < \epsilon$,
which is used to make the peak-to-average power ratio (PAPR) of one user within an acceptable range.


\subsection{Construction}

There are many approaches to construct a $T$-size UDAS set. This paper presents two of them, which are
cyclic and quasi-cyclic (QC) matrix based construction schemes.

\subsubsection{Cyclic matrix mode}
Let ${\bf E}_{cyc}$ be a cyclic matrix, and the subscript ``cyc'' indicates ``cyclic''. The first row of ${\bf E}_{cyc}$ is viewed as a generator of ${\bf E}_{cyc}$, denoted by a $1 \times T$ vector, i.e., ${\bf a} = (a_1, a_2, \ldots, a_t, \ldots, a_T)$, where $a_t$ is an element in $\Delta$. Each element of ${\bf a}$ is shifted to the rightward and downward one position, and the last element of ${\bf a}$ is moved to the first position, achieving the second row of ${\bf E}_{cyc}$. Repeat this processing, until the first element $a_1$ of ${\bf a}$ has researched the last position of the $T$th row. ${\bf E}_{cyc}$ is a $T \times T$ matrix, given by
\begin{equation}\label{e_cyc}
{\bf E}_{cyc} =
\left[
	\begin{matrix}
		{\bf e}_1 \\
		{\bf e}_2 \\
		\vdots \\
		{\bf e}_T \\
	\end{matrix}
\right]
	=
\left[
	\begin{matrix}
		a_{1}  & a_{2}  & \ldots  & a_{T} \\
		a_{T}  & a_{1}  & \ldots  & a_{T-1} \\
		\vdots & \vdots & \ddots  & \vdots  \\
		a_{2}  & a_{3}  & \ldots  & a_{1} \\
	\end{matrix}
\right].
\end{equation}

Because of the cyclic structure, the rows of ${\bf E}_{cyc}$, i.e., ${\bf e}_1, {\bf e}_2,\ldots,{\bf e}_T$, are formed a $T$-size UDAS set $\Psi$. Moreover, it is easy to derive that 
$P_{avg} = P_{t} = \frac{1}{L} \sum_{l=1}^{L} a_{l}^2$, where $1 \le t \le T$ and $L=T$.
We give an example to show the construct processing.

\textbf{Example 2:}
Assume $L = 4$, and the generator ${\bf a}= (1, 1i, 2, 2i )$. Then,
\begin{equation} \label{e_Ex3}
{\bf E}_{cyc} =
\left[
	\begin{matrix}
		{\bf e}_1 \\
		{\bf e}_2 \\
		{\bf e}_3 \\
		{\bf e}_4 \\
	\end{matrix}
\right] = \left[
\begin{matrix}
		1 & 1i & 2 & 2i \\
		2i & 1 & 1i & 2 \\
		2 & 2i & 1 & 1i \\
		1i & 2 & 2i & 1 \\
	\end{matrix}
	\right],
\end{equation}
The average power $P_{avg}$ equals to $2.5$. End Example 2.

Recall Example 1, we have obtained ${\bf c}_1^{(j)} = \{011 \quad 100 \quad 001 \quad 110\}$
and ${\bf c}_2^{(j)} = \{100 \quad 010 \\ \quad 111 \quad 001 \}$.
Assume `00', `01', `10' and `11' are respectively mapped to four different locations, suppose ${\bf e}_j = (1, 1i, 2, 2i)$, it is able to obtain
\begin{equation*}
	\begin{matrix}
    x_{1,1}^{(j)} = (0, +1, 0, 0), &
    x_{1,2}^{(j)} = (0, 0, -1i, 0), &
    x_{1,3}^{(j)} = (+2, 0, 0, 0), &
    x_{1,4}^{(j)} = (0, 0, 0, -2i), \\
    x_{2,1}^{(j)} = (0, 0, -1, 0), &
    x_{2,2}^{(j)} = (0, -1i, 0, 0), &
    x_{2,3}^{(j)} = (0, 0, 0, +2), &
    x_{2,4}^{(j)} = (+2i, 0, 0, 0), \\
	\end{matrix}
\end{equation*}
and the modulated ${\bf x}_{1}^{(j)}$ and ${\bf x}_{2}^{(j)}$ are transmitted to the channel.

\subsubsection{Quasi-cyclic matrix mode}
A $T$-size UDAS set with quasi-cyclic (QC) structure is defined by ${\bf E}_{qc}$, where the subscript ``qc'' stands for ``quasi-cyclic''. Refer to the QC structure of QC-LDPC codes \cite{LinBook3}, 
let ${\bf E}_{qc}$ be an $S \times Q$ array, and each position of ${\bf E}_{qc}$ is corresponding to a matrix, as
\begin{equation} \label{e_qc}
{\bf E}_{qc} =
\left[
	\begin{matrix}
		{\bf A}_{1,1} & {\bf A}_{1,2} & \ldots & {\bf A}_{1,Q}\\
		{\bf A}_{2,1} & {\bf A}_{2,2} & \ldots & {\bf A}_{2,Q}\\
		\vdots & \vdots & \ddots & \vdots\\
		{\bf A}_{S,1} & {\bf A}_{S,2} & \ldots & {\bf A}_{S,Q}\\
	\end{matrix}
\right],
\end{equation}
where ${\bf A}_{s,q}$ is an ${\mathcal L} \times {\mathcal L}$ cyclic matrix.
Hence, ${\bf E}_{qc}$ is an $({\mathcal L} \cdot S) \times ({\mathcal L} \cdot Q)$ quasi-cyclic matrix.
Denote the generator of ${\bf A}_{s,q}$ by ${\bf a}_{s,q}$, i.e., ${\bf a}_{s,q} = (a_{s,q,1}, a_{s,q,2},\ldots, a_{s,q, \iota}, \ldots, a_{s,q, \mathcal L} )$,
where $1 \le s \le S$, $1 \le q \le Q$ and $1 \le \iota \le \mathcal L$.
Given $q$ and $\iota$, if $s \neq s'$, ${\bf a}_{s,q,\iota} \neq {\bf a}_{s',q,\iota}$,
then the rows of ${\bf E}_{qc}$ can form an $({\mathcal L} \cdot S)$-size UDAS set,
in which each sequence is with length $L = {\mathcal L} \cdot Q$.

Equation (\ref{e_qc}) is a general expression of the QC structure, and it can be extended to a block-wise cyclic structure, given by
\begin{equation} \label{e_bqc}
{\bf E}_{b,qc} =
\left[
	\begin{matrix}
		{\bf A}_{1,1} & {\bf A}_{1,2} & \ldots & {\bf A}_{1,Q}\\
		{\bf A}_{1,Q} & {\bf A}_{1,1} & \ldots & {\bf A}_{1,Q-1}\\
		\vdots & \vdots & \ddots & \vdots\\
		{\bf A}_{1,2} & {\bf A}_{1,3} & \ldots & {\bf A}_{1,1}\\
	\end{matrix}
\right],
\end{equation}
where ${\bf A}_{1,q}$ is an ${\mathcal L} \times {\mathcal L}$ cyclic matrix.

It is seen that the first row-block is shifted to the rightward and downward one position and the last block is moved to the first position, leading to a new row-block. Each row-block of (\ref{e_bqc}) has cyclic structure, thus ${\bf E}_{b,qc}$ is an $({\mathcal L} \cdot Q) \times ({\mathcal L} \cdot Q)$ block-wise cyclic square matrix, where ``$b$'' stands for ``block-wise''.

\textbf{Example 3:}
Let $Q = 3$, and the generators of ${\bf A}_{1,1}, {\bf A}_{1,2}$ and ${\bf A}_{1,3}$ are respectively
${\bf a}_{1,1}= (1, 1i), {\bf a}_{1,2}= (2, 2i)$ and ${\bf a}_{1,3}= (4, 4i)$.
Then, the block-wise QC structure ${\bf E}_{b,qc}$ is given by
\begin{equation} \label{e_exam4}
{\bf E}_{b,qc} = \left[
	\begin{matrix}
		1 & 1i & 2 & 2i & 4 & 4i\\
		1i & 1 & 2i & 2 & 4i & 4\\
		4 & 4i & 1 & 1i & 2 & 2i\\
		4i & 4 & 1i & 1 & 2i & 2\\
		2 & 2i & 4 & 4i & 1 & 1i\\
		2i & 2 & 4i & 4 & 1i & 1\\
	\end{matrix}
\right].
\end{equation}
Each row of ${\bf E}_{b,qc}$ is an UDAS, with average power $P_{avg} = 7$. End Example 3.

\subsection{Features of the proposed cyclic/quasi-cyclic UDAS set}
As aforementioned definition, the sum-pattern vector set of $T$-size UDAS set 
$\Psi$ is $\Omega^T = \{ \Omega_{1}^T, \Omega_{2}^T, \ldots, \Omega_{l}^T,\ldots, \Omega_{L}^T\}$.
However, for a random multiple access scenario, not all the $T$ users are simultaneously transmission.
Now, we consider a general case.

Assume $\tau$ ($1 \le \tau \le T$) different sequences of $\Psi$ are selected for multiuser transmission, 
i.e., $\Psi^{(\tau, \mu)} = \{ {\bf e}_{t_1}, \ldots, {\bf e}_{t_\nu}, \ldots, {\bf e}_{t_\tau} \}$, 
where $t_\nu \in \{1, 2, \ldots, T\}$ and $\mu$ is the  $\tau$-size combination index.
Actually, there are totally $C_T^{\tau}$ combinations, indicating $1 \le \mu \le C_T^{\tau}$.
At this moment, the sum-pattern of the $l$th symbol of the $\mu$th combination index is
$w_l = \sum_{\nu = 1}^{\tau} (2c_l^{(\nu)} - 1) \cdot e_{t_\nu,l}$,
where $w_l \in \omega_l^{(\tau,\mu)}$, and the number of elements in $\omega_l^{(\tau,\mu)}$ equals to
$|\omega_l^{(\tau,\mu)}| = 2^{\tau}$.

Define the sum-pattern vector set of $\tau$ users by $\Omega^{\tau} = \{ \Omega_{1}^{\tau}, \Omega_{2}^{\tau}, ..., \Omega_{l}^{\tau},\ldots, \Omega_{L}^{\tau}\}$, where $\Omega_l^{\tau} = \omega_l^{(\tau,1)} \cup \omega_l^{(\tau,2)} \cup \ldots \omega_l^{(\tau,\mu)}\ldots \cup \omega_l^{(\tau, C_T^{\tau})}$ 
and the number of elements in $\Omega_{l}^{\tau}$ satisfies $|\Omega_{l}^{\tau}| \le C_T^{\tau} \cdot 2^{\tau}$.
When $\tau = T$, it is found that $|\Omega_{l}^{T}| = 2^T$, equaling to the $T$-size case.

Moreover, define the maximum values of the real and image parts of $\omega_l^{(\tau,\mu)}$ by
\begin{equation*}
	\begin{aligned}
		\kappa_{l,re}^{(\tau,\mu)} &= {\rm max} \left\{ {\rm Re}[\omega_l^{(\tau,\mu)}] \right\},
		\kappa_{l,im}^{(\tau,\mu)} &= {\rm max} \left\{ {\rm Im}[\omega_l^{(\tau,\mu)}] \right\}.
	\end{aligned}
\end{equation*}
While, $(\kappa_{l,re}^{(\tau,\mu)}, \kappa_{l,im}^{(\tau,\mu)})$ are used for finding $\omega_l^{(\tau,\mu)}$.
The average power of the sum-patterns in $\omega_l^{(\tau,\mu)}$, defined by $\lambda_{l}^{(\tau,\mu)}$, is equal to
\begin{equation*}
	\begin{matrix}
		\lambda_{l}^{(\tau,\mu)} 
		=\frac{1}{2^\tau} \sum_{w_l \in \omega_l^{(\tau,\mu)}} \|w_{l}\|^2 
		= \sum_{\nu = 1}^{\tau} e_{t_\nu,l}^2,
	\end{matrix}
\end{equation*}
relaying on the values of $\{e_{t_1,l}, e_{t_2,l},\ldots, e_{t_\nu,l}, \ldots, e_{t_{\tau},l}\}$.
For a given $\tau$, define the average power of $\omega_l^{(\tau,\mu)}$ of the total $L$ locations by
$\Lambda^{(\tau,\mu)} = \{\lambda_1^{(\tau,\mu)}, \lambda_2^{(\tau,\mu)},\ldots, \lambda_l^{(\tau,\mu)},\ldots,
\lambda_L^{(\tau,\mu)}\}$,
assisting AUD and multiuser detection.
It is easy to derive that the sum of $\lambda_{l}^{(\tau,\mu)}$ is
\begin{equation*}
	\begin{aligned}
		\lambda_{sum}^{\tau} &= \sum_{l=1}^L \lambda_{l}^{(\tau,\mu)}
		                        = \sum_{\nu = 1}^{\tau} \sum_{l=1}^{L} e_{t_\nu,l}^2
							    = \tau \cdot L P_{avg},
	\end{aligned}
\end{equation*}
which is a constant for a given $\tau$, independent of the selection of sequences, i.e., combination index $\mu$.
This interesting result can help us quickly detect the number of arrival users.
To illustrate this result, let us have a look at another example.

\begin{figure}[t]
  \centering
  \includegraphics[width=0.9\linewidth]{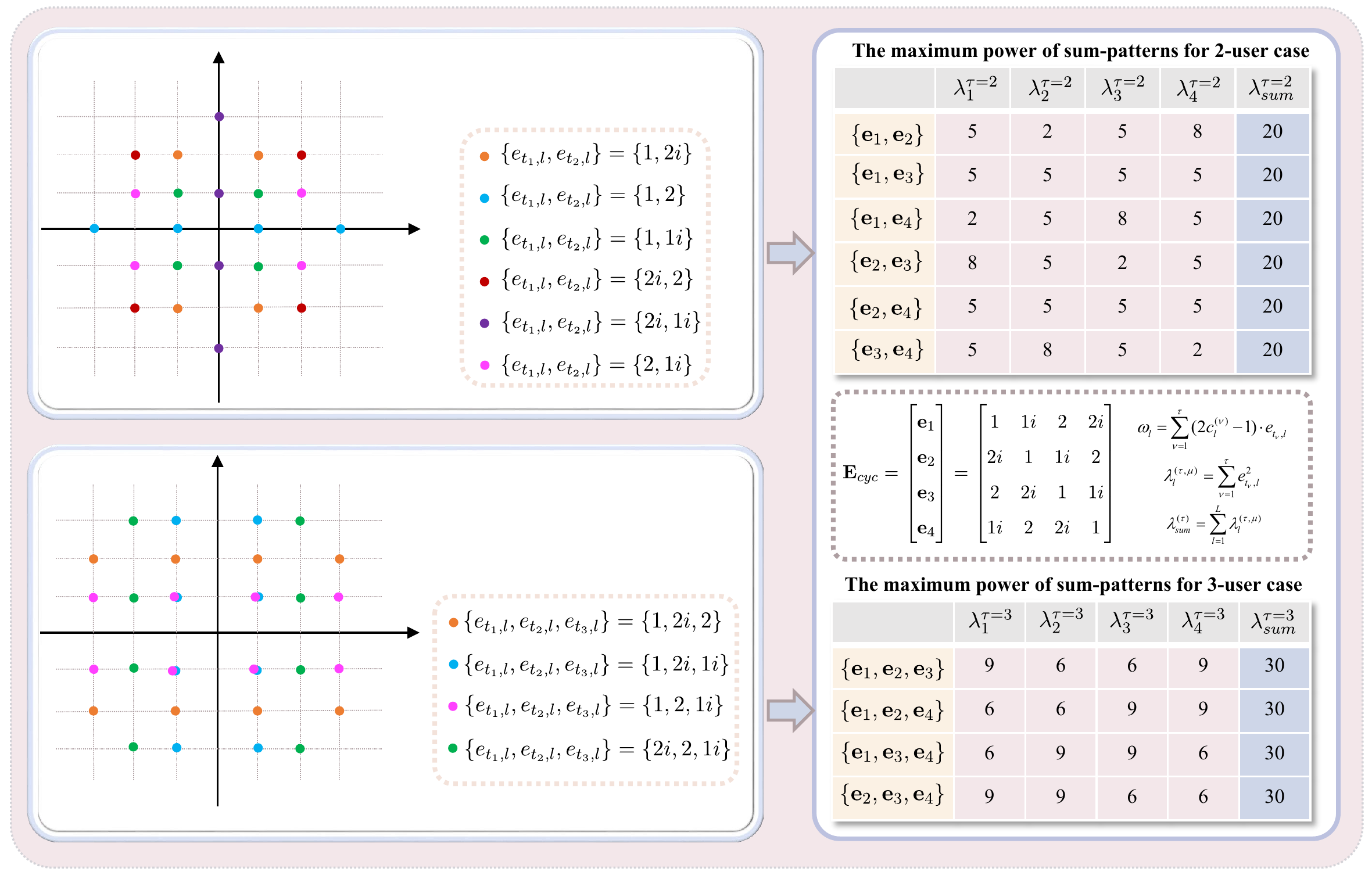}
  \caption{An illustration of Example 4, where $\Psi =  \{ {\bf e}_1, {\bf e}_2, {\bf e}_3, {\bf e}_4 \} = \{(1, 1i, 2, 2i), (2i, 1, 1i, 2), (2, 2i, 1, 1i), (1i, 2, 2i, 1)\}$. The upper part of the figure is for the case of $\tau=2$, and the lower part of the figure is for the case of $\tau=3$.}
  \label{Fig3}
\end{figure}
  
\textbf{Example 4:}
Recall (\ref{e_Ex3}) of Example 2, there are four sequences,
$\Psi =  \{ {\bf e}_1, {\bf e}_2, {\bf e}_3, {\bf e}_4 \} = \{(1, 1i, 2, 2i), (2i, 1, 1i, 2), (2, 2i, 1, 1i), (1i, 2, 2i, 1)\}$, with $P_{avg} = 2.5$ and $L \cdot P_{avg} = 10$.

If we select any $\tau = 2$ sequences from the given $\Psi$ for multiuser transmission, there are totally $C_4^2 = 6$ combinations, i.e.,
$\Psi^{(2,1)} = \{ {\bf e}_1, {\bf e}_2\}$, $\Psi^{(2,2)} = \{ {\bf e}_1, {\bf e}_3\}$,
$\Psi^{(2,3)} = \{ {\bf e}_1, {\bf e}_4\}$, $\Psi^{(2,4)} = \{ {\bf e}_2, {\bf e}_3\}$,
$\Psi^{(2,5)} = \{ {\bf e}_2, {\bf e}_4\}$, and $\Psi^{(2,6)} = \{ {\bf e}_3, {\bf e}_4\}$.
The size of the sum-pattern set $\omega_l^{(\tau,\mu)}$ is $|\omega_l^{(\tau,\mu)}| = 2^2 = 4$, 
where $1 \le \mu \le C_4^2$. When $l=1$, we have
\begin{small}
\begin{equation*}
	\begin{aligned}
	\{1, 2i\} & \rightarrow \omega_{1}^{(2,1)} = \{1+2i, 1-2i, -1+2i, -1-2i\} \rightarrow \lambda_{1}^{(2,1)} = 5;\\
  \{1, 2\} & \rightarrow \omega_{1}^{(2,2)} = \{-3, -1, 1, 3\} \rightarrow \lambda_{1}^{(2,2)} = 5;\\
    \{1, 1i\} & \rightarrow \omega_{1}^{(2,3)} = \{1+1i, 1-1i, -1+1i, -1-1i\} \rightarrow \lambda_{1}^{(2,3)} = 2;\\
    \{2i, 2\} & \rightarrow \omega_{1}^{(2,4)} = \{2+2i, 2-2i, -2+2i, -2-2i\} \rightarrow \lambda_{1}^{(2,4)} = 8;\\
    \{2i, 1i\} & \rightarrow \omega_{1}^{(2,5)} = \{-3i, -1i, 1i, 3i \}  \rightarrow \lambda_{1}^{(2,5)} = 5;\\
    \{2, 1i\} & \rightarrow \omega_{1}^{(2,6)} = \{2+1i, 2-1i, -2+1i, -2-1i \} \rightarrow \lambda_{1}^{(2,6)} = 5.
  \end{aligned}
\end{equation*}
\end{small}
It is found that $\omega_l^{(2,\mu)} \cap \omega_l^{(2,\mu')} = \emptyset$, when $\mu \neq \mu'$.
Moreover, it is able to derive that $\lambda_{sum}^{\tau=2} = 20$, which equals to $2 \times (L \cdot P_{avg})$.

Observe $\Lambda^{(2, \mu)}$, it is found that $\Lambda^{(2, 2)} = \Lambda^{(2, 5)}$, i.e., $(5, 5, 5, 5)$; thus, $\Psi^{(2,2)}$ and $\Psi^{(2,5)}$ cannot be separated based on their $\Lambda$.
However, observe $\omega_1^{(2,2)}$ and $\omega_1^{(2,5)}$ again, it is found that
$(\kappa_{l,re}^{(2,2)}, \kappa_{l,im}^{(2,2)}) = (3, 0)$ and
$(\kappa_{l,re}^{(2,5)}, \kappa_{l,im}^{(2,5)}) = (0, 3)$,
indicating that $\Psi^{(2,2)}$ and $\Psi^{(2,5)}$ can be separated based on their corresponding $(\kappa_{l,re}^{(\tau,\mu)}, \kappa_{l,im}^{(\tau,\mu)})$.

If we pick any $\tau = 3$ sequences from $\Psi$ for multiuser transmission, 
there are $C_4^3 = 4$ combinations, i.e.,
$\Psi^{(3,1)} = \{ {\bf e}_1, {\bf e}_2, {\bf e}_3\}$,
$\Psi^{(3,2)} = \{ {\bf e}_1, {\bf e}_2, {\bf e}_4\}$,
$\Psi^{(3,3)} = \{ {\bf e}_1, {\bf e}_3, {\bf e}_4\}$, and $\Psi^{(3,4)} = \{ {\bf e}_2, {\bf e}_3, {\bf e}_4\}$.
At this time, the sum-pattern sets of the $\mu$th combination of the $l$th symbol can be given. 
Take $l=1$ as an example, they are
\begin{small}
\begin{equation*}
	\begin{aligned}
	\{1, 2i, 2\} \rightarrow \omega_{1}^{(3,1)} = \{-3+2i, -3-2i, -1+2i, -1-2i, 1+2i, 1-2i, 3+2i, 3-2i \}
	\rightarrow \lambda_{1}^{(3,1)} = 9;\\
	\{1, 2i, 1i\} \rightarrow \omega_{1}^{(3,2)} = \{ 1-3i, -1-3i, 1-1i, -1-1i, 1+1i, -1+1i, 1+3i, -1-3i \} \rightarrow \lambda_{1}^{(3,2)} = 6;\\
	\{1, 2, 1i\}\rightarrow \omega_{1}^{(3,3)} = \{-3+1i, -3-1i, -1+1i, -1-1i, 1+1i, 1-1i, 3+1i, 3-1i \}
	\rightarrow \lambda_{1}^{(3,3)} = 6;\\
	\{2i, 2, 1i\}\rightarrow \omega_{1}^{(3,4)} = \{ 2-3i, -2-3i, 2-1i, -2-1i,2+1i, -2+1i, 2+3i, -2+3i \}
	\rightarrow \lambda_{1}^{(3,4)} = 9.\\
	\end{aligned}
\end{equation*}
\end{small}
For this scenario, we can know $\lambda_{sum}^{\tau=3} = 30$ that is equal to $3 \times (L \cdot P_{avg})$.
Moreover, it is found that $\omega_l^{(3,2)} \cap \omega_l^{(3,3)} = \{1-1i, -1-1i, -1+1i, -1-1i \}$, indicating that there exists constellation overlap.
Observe the four sets $\Lambda^{(3,1)}, \Lambda^{(3,2)}, \Lambda^{(3,3)}$ and $\Lambda^{(3,4)}$, they are different from each other, thus it is a one-to-one mapping between $\Lambda^{(3,\mu)}$ and $\Psi^{(3,\mu)}$, i.e., $\Lambda^{(\tau,\mu)} \to \Psi^{(\tau,\mu)}$.
For this scenario, we can calculate $\Lambda^{(\tau, \mu)}$, 
and then obtain the corresponding $\Psi^{(\tau,\mu)}$. End Example 4.

\section{Active user detection and multiuser detection}
This section presents a statistic of UDAS feature based active user detection (SoF-AUD), 
and a joint multiuser detection (MUD) and improved iterative message passing algorithm (MPA) for the proposed system.
Assume that the users are assigned different sequences from the $T$-size UDAS set ${\Psi}$.
The receiver will detect the number of arrival users and recover their individual data information.

\subsection{MAP detection}
First of all, we introduce the maximum {\it a posterior} (MAP) detection algorithm, which is an optimal solution for the proposed system.
Take (\ref{e_s_mlj}) into (\ref{e4}), $y_{m,l,i}$ is rewritten as
\begin{equation} \label{e_r_iml}
	\begin{split}
y_{m,l,i} = \sum_{j=1}^{J} \rho_{m,l,i}^{(j)} \cdot \left( 2{c}_{m,l,\log_2{\mathcal M} }^{(j)}-1 \right) 
			\cdot {e}_{j,l} + z_{m,l,i} = w_{m,l,i} + z_{m,l,i},
	\end{split}
\end{equation}
where $w_{m,l,i} = \sum_{j=1}^{J} \rho_{m,l,i}^{(j)} \cdot \left( 2{c}_{m, l, \log_2{\mathcal M} }^{(j)}-1 \right) \cdot {e}_{j,l}$ is the received superimposed signal (or called sum-pattern). 
For further discussion, set ${\bf w} = ({\bf w}_1, {\bf w}_2, \ldots, {\bf w}_m, \ldots, {\bf w}_M)$, 
and
${\bf w}_m = ( {\bf w}_{m,1}, {\bf w}_{m,2}, \ldots, {\bf w}_{m,l},\ldots, {\bf w}_{m,L} )$, 
where
${\bf w}_{m,l} = (w_{m,l,1}, w_{m,l,2},\ldots, w_{m,l,i},\ldots, w_{m,l, \mathcal M_1})$.

Equation (\ref{e_r_iml}) includes variables $J$, $\rho_{m,l,i}^{(j)}$, ${c}_{m, l,\log_2{\mathcal M} }^{(j)}$ and ${e}_{j,l}$.
Actually, $\rho_{m,l,i}^{(j)}$ and ${c}_{m, l, \log_2{\mathcal M} }^{(j)}$ together are 
corresponding to ${\bf c}_{m,l}^{(j)}$. 
Based on the MAP criterion, it is able to derive that
\begin{equation}
\left(J, {\bf C}, \Psi^{(J,\mu)}  \right) =
 {\rm argmax} \left[P_{J} \cdot \exp \left\{- \frac{\left\| {\bf y} - {\bf w}  \right\|^2}{N_0} \right\}  \right],
\end{equation}
where ${\bf C} =\{ {\bf C}^{(1)},\ldots,{\bf C}^{(J)}\}$ is the set of transmit bits of all $J$ users, $\Psi^{(J,\mu)}$ is the selected UDAS set, and $P_{J}$ is probability of $J$-user simultaneously access the receiver. 
Assume the receiver can maximum detect $J_{max}$ users. 
When $J > J_{max}$, the detection is interrupted and all the users' packets are lost.
Considering the application of the $T$-size UDAS set $\Psi$, 
in the following discussion, we assume $J_{max} = T$ and $J = \tau$.

Although the MAP detection is the optimal algorithm, the complexity is extremely high for the proposed system,
i.e., $\mathcal{O}{\left( \left( L \cdot  {\mathcal M}^{ML}   \right)^T\right)}$.
Thereby, regarding as the features of the cyclic/quasi-cyclic UDAS set, 
we firstly introduce a SoF-AUD algorithm, whose complexity is extremely low.

\subsection{Step 1: SoF-AUD}

The SoF-AUD includes two parts, one is to detect the number of arrival users $\tau$ ($1 \le \tau \le T$), and the other is to find the selected UDAS set ${\Psi}^{(\tau,\mu)}$.

For the $l$th received symbol of ${\bf y}_m$, 
define the $i$th location of the $l$th received symbol of all $M$ rows by
${\bf y}_{l,i}^{\rm{loc}} = \left(y_{1,l,i}, y_{2,l,i}, \ldots, y_{m,l,i}, \ldots, y_{M,l,i} \right)$.
If we ignore the effect of noise and $M$ is large enough, ${\bf y}_{l,i}^{\rm loc}$ can traverse all the constellations of $\omega_l^{(\tau,\mu)}$, including $(\kappa_{l,re}^{(\tau,\mu)}, \kappa_{l,im}^{(\tau,\mu)})$.
Actually, the proposed SoF-AUD algorithm is relayed on the statistical properties of ${\bf y}_{l,i}^{\rm loc}$.

The sum power $\gamma_{sum}$ of the entire $L$ symbols can be calculated as
\begin{equation} \label{e_gamma}
	\begin{aligned}
		\gamma_{sum} &= \sum_{l=1}^{L} \gamma_{l}
							= \sum_{l=1}^{L} \left\{ \frac{1}{M}\sum_{m=1}^{M} \sum_{i=1}^{\mathcal{M}_1}  
							\left\| y_{m,l,i} \right\|^2 \right\},
	\end{aligned}
\end{equation}
where $\gamma_{l} =\frac{1}{M} \sum_{m = 1}^{M} \left[ \sum_{i=1}^{\mathcal{M}_1} \|y_{m,l,i}\|^2 \right]$ is the average power of the $l$th symbol.

Thus, for a given number of arrival users $\tau$, the statistical mean of $\gamma_{sum}$ can be deduced as
\begin{equation}\label{E_gammaSum}
	\begin{aligned}
\overline \gamma_{sum}^{\tau} 
	&= {\mathbb E} \left[  \sum_{l=1}^{L} \left\{ \frac{1}{M}\sum_{m=1}^{M} \sum_{i=1}^{\mathcal{M}_1}  
							\left\| y_{m,l,i} \right\|^2 \right\} \right] 
							= {\mathbb E} \left[  \sum_{l=1}^{L} \sum_{i=1}^{\mathcal{M}_1}
							\left\| w_{m,l,i} + z_{m,l,i}\right\|^2 \right] \\
	&=  \sum_{l=1}^{L} \sum_{i=1}^{\mathcal{M}_1} {\mathbb E} \left[ 
							\left\| w_{m,l,i} + z_{m,l,i}\right\|^2 \right] = \lambda_{sum}^{\tau} + L \mathcal{M}_1 \cdot N_0
	\end{aligned},
\end{equation}
since ${\mathbb E} \left[ w_{m,l,i} \cdot z_{m,l,i}  \right] = 0$ 
and ${\mathbb E} \left[\|z_{m,l,i}\|^2 \right] = N_0$.
Equation (\ref{E_gammaSum}) is the sum of $\lambda_{sum}^{\tau}$ and $L \mathcal{M}_1 \cdot N_0$,
indicating $\overline \gamma_{sum}^{\tau}$ is a constant for a given $\tau$.
Here, $\lambda_{sum}^{\tau} = {\tau} \cdot (L P_{avg})$, and $L P_{avg}$ is a constant for a given cyclic/quasi-cyclic UDAS set $\Psi$.

Since we utilize $\mathcal M_1$-dimensional modulation, 
the constellation with the maximum power of the $l$th symbol in ${\bf y}_{l,i}^{\rm loc}$ can be calculated as
\begin{equation*}
	\begin{aligned}
		\zeta_{l,re} &= \sum_{i=1}^{\mathcal M_1} \left[ \max_{1 \le m \le M} \left\{ {\rm Re}[{\bf y}_{l,i}^{\rm loc}] \right\} \right], \qquad
		\zeta_{l,im} &= \sum_{i=1}^{\mathcal M_1} \left[ \max_{1 \le m \le M} \left\{ {\rm Im}[{\bf y}_{l,i}^{\rm loc}] \right\} \right].
	\end{aligned}
\end{equation*}

At this time, we can estimate the number of arrival users based on minimum square error (MSE) criterion as
\begin{equation} \label{e_L1andL2}
	\begin{aligned}
		({\widehat \tau, \widetilde{\mu}}) 
		&= \arg\min_{1 \le \tau \le T} 
		   \left\|\gamma_{sum} - \overline \gamma_{sum}^{\tau} \right\| = \arg\min_{1 \le \tau \le T} 
	       \left\|\gamma_{sum} - (\lambda_{sum}^{\tau} + L \mathcal{M}_1 \cdot N_0)\right\|, \\
      s.t. \quad &C1:
	    {\widehat \mu} = \arg\min_{\mu' = \widetilde{\mu}} \sum_{l=1}^{L} d_{l}^{(\tau, \mu')} , \\
	\end{aligned}
\end{equation}
where $d_{l}^{(\tau, \mu')} = \|\zeta_{l,re} - \kappa_{l,re}^{(\tau,\mu')}\|_2 + \|\zeta_{l,im} - \kappa_{l,im}^{(\tau,\mu')}\|_2$.
Consider there are more than one $\widetilde \mu$ can satisfy (\ref{e_L1andL2}), $C1$ is used to find the only one UDAS set.

According to (\ref{e_L1andL2}), it is able to obtain the number of arrival users $\widehat \tau$ and the selected UDAS set $\Psi^{(\widehat \tau, \widehat \mu)}$. The complexity of the AUD modular is $ \mathcal{O}(ML^2)$.
\vspace{-15pt}

\subsection{Step 2: MUD}
Based on the $\widehat \tau$ and $\Psi^{(\widehat \tau, \widehat \mu)}$ obtained by SoF-AUD, we begin to detect and separate the superimposed signals.
Actually, we have taken into consideration the connection (or statistics) among the received signals of $M$ rows
(i.e.,${\bf y}_1, {\bf y}_2, \ldots, {\bf y}_M$)
during the AUD process. Now, we can deal with the received signals row-by-row to reduce the complexity.

\subsubsection{${\mathcal M}_1 = 1$}
When ${\mathcal M}_1 = 1$, ${x}_{m,l}^{(j)}$ is equivalent to ${s}_{m,l}^{(j)}$. 
The detected results ${\widehat w}_{m,l,1}$ is selected from the set $\Psi^{({\widehat \tau}, \widehat \mu)}$, expressed as
\begin{equation} \label{e_MUD_hard}
	\begin{aligned}
	{\widehat \omega}_{m,l,1} &= \arg\min_{\omega_{m,l,1} \in \Psi^{({\widehat \tau}, \widehat \mu)}}
	\| y_{m,l,1} - \omega_{m,l,1} \|_2, \\
	s.t. \quad &C2:
	\Psi^{(\widehat \tau, \widehat \mu)} \to {\widehat \omega}_{m,l,1} \to 
	\{{c_{m,l,1}^{(1)}},  {c_{m,l,1}^{(2)}}, \ldots, {c_{m,l,1}^{(\widehat \tau)}}  \},\\
	&C3:
	0 = {c_{m,l,1}^{(j)}} \oplus {c_{m,2,1}^{(j)}} \oplus \ldots \oplus {c_{m,L,1}^{(j)}}.
	\end{aligned}
\end{equation}
$C2$ stands for the one-to-one mapping between ${\widehat \omega}_{m,l,1}$ 
and $\{{c_{m,l,1}^{(1)}},  {c_{m,l,1}^{(2)}}, \ldots, {c_{m,l,1}^{(\widehat \tau)}}  \}$. 
$C3$ reflects the SPC constraint, where $1\le j \le \widehat \tau$.

It is remarkable that (\ref{e_MUD_hard}) is a hard decision processing, which helps fast realize MUD with acceptable BER. It can be applied to the scenario without ${\bf G}_1$.

\subsubsection{${\mathcal M}_1 > 1$}
When ${\mathcal M}_1 > 1$, the detection processing becomes a little complex.  
Define the sum-pattern set of the $i$th location of the $\mathcal M_1$-dimensional modulation by $\Psi^{(\widehat \tau_i, \widehat \mu_i)}$, where $\widehat \tau_i$ and $\widehat \mu_i$ are respectively $\widehat \tau_i$ users and the $\widehat \mu_i$th combination at the $i$th location. 
Let the set of $\mathcal M_1$-dimensional sum-pattern be 
${\Theta}^{({\widehat \tau}, \widehat \mu)} = ( \Psi^{(\widehat \tau_1,  \widehat \mu_1)}, \Psi^{(\widehat \tau_2, \widehat \mu_2)} , \ldots, \Psi^{ (\widehat \tau_i,  \widehat \mu_i)} , \ldots, \Psi^{ (\widehat \tau_{\mathcal{M}_1},  \widehat \mu_{\mathcal{M}_1})} )$, the detected results are then expressed as
\begin{equation} \label{e_MUD_FSK_hard}
	\begin{aligned}
	{\widehat \omega}_{m,l}
	    &= \arg\min_{\theta_{m,l} \in {\Theta}^{({\widehat \tau}, \widehat \mu)}}
	       \| y_{m,l} - {\theta_{m,l}} \|_2, \\
	s.t. \quad  &C4: 
  		\widehat {\tau} = \sum_{i=1}^{\mathcal{M}_1} \widehat \tau_i, \\
  	 &C5: 
  		\mu_i \subseteq \mu,\\
   	 &C6: 
		{\widehat \omega}_{m,l} \to \{{\bf c}_{m,l}^{(1)} , {\bf c}_{m,l}^{(2)}, \ldots, {\bf c}_{m,l}^{(\widehat \tau)} \},\\
	 &C7: 
		0 = c_{m,1,\log_2{\mathcal M}}^{(j)} \oplus c_{m,2,\log_2{\mathcal M}}^{(j)} 
		\oplus \ldots \oplus c_{m,L,\log_2{\mathcal M}}^{(j)}.
	\end{aligned}
\end{equation}
where ${\widehat \omega}_{m,l}=\left({\widehat \omega}_{m,l,1}, {\widehat \omega}_{m,l,2}, \ldots, {\widehat \omega}_{m,l,\mathcal M_1}\right)$,
$\theta_{m,l}=(\omega_{m,l,1}, \omega_{m,l,2}, \ldots, \omega_{m,l,\mathcal M_1})$ and 
$\omega_{m,l,i} \in \Psi^{(\widehat \tau_i, \widehat \mu_i)}$ for $1 \le i \le \mathcal M_1$.
$C4$ is to keep the sum of the detected number of users be a constant.
$C5$ indicates that the users are from the same selected UDAS set.
$C6$ shows the one-to-one mapping between ${\widehat \omega}_{m,l}$ and $\{{\bf c}_{m,l}^{(1)} , {\bf c}_{m,l}^{(2)}, \ldots, {\bf c}_{m,l}^{(\widehat \tau)} \}$. 
$C7$ also stands for the SPC constraint of ${\bf G}_2$, where $1 \le j \le \widehat{\tau}$.

Actually, (\ref{e_MUD_FSK_hard}) can be viewed as an extension of (\ref{e_MUD_hard}) to multi-dimensional case.
The mainly difference is that the referred set is from $\Psi^{(\tau, \mu)}$ to ${\Theta}^{(\tau, \mu)}$,
with complexity $\mathcal{O} \left( \mathcal{M}^{ML \cdot \widehat \tau} \right)$.

\subsection{Step 3: Improved MPA}

When a LDPC code is utilized as the first encoder ${\bf G}_1$, it is important to find the initial LLR (log likelihood ratio) according to the received signals for further message passing decoding. In other words, we should calculate the LLRs from the MUD process.

\begin{figure}[tbp]
  \centering
  \includegraphics[width=0.6\linewidth]{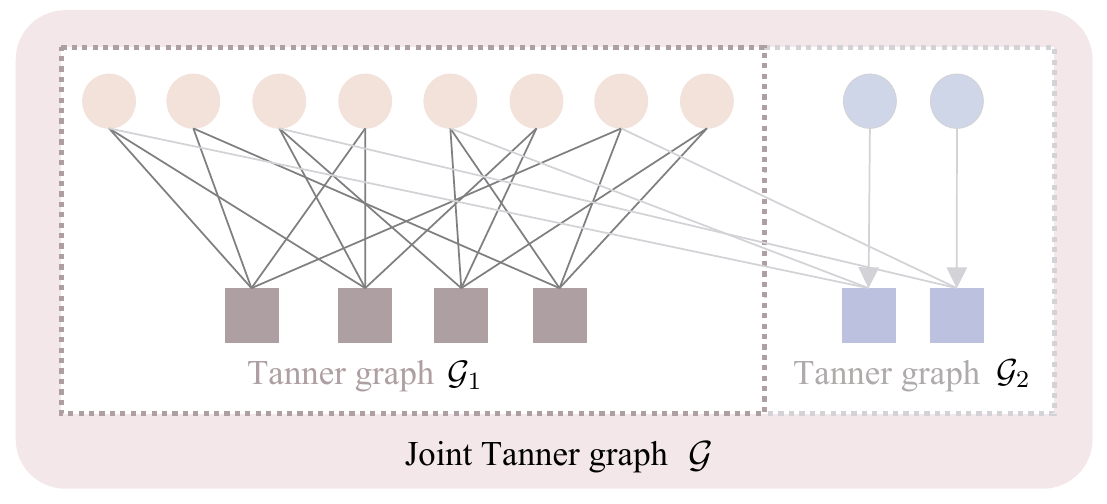}
  \caption{A joint Tanner graph of the two encoders $\mathcal G$, where $\mathcal G_1$ and $\mathcal G_2$ are Tanner graphs of encoder one and encoder two, respectively.}
  \label{Fig4}
  \end{figure}

\subsubsection{Initial LLRs}
The probability of ${\widehat \omega}_{m,l}$ can be calculated as
\begin{equation} \label{e_Pro_fre}
	P({\widehat \omega}_{m,l}) = \frac{1}{\sqrt{\pi N_0}} {\rm exp} 
  		\left\{- \frac{\| y_{m,l} - {\widehat \omega}_{m,l} \|^2}{N_0} \right\},
\end{equation}
where ${\widehat \omega}_{m,l} \in {\Theta}^{({\widehat \tau}, \widehat \mu)}$. 
Since ${\widehat \omega}_{m,l}$ is corresponding to $\log_2{\mathcal{M}}$ bits of $\widehat \tau$ users, i.e.,
\begin{equation*}
\begin{aligned}
{\widehat \omega}_{m,l} \to 
		\{{\bf c}_{m,l}^{(1)} , {\bf c}_{m,l}^{(2)}, \ldots, {\bf c}_{m,l}^{(\widehat \tau)} \} 
		&= \{ (c_{m,l,1}^{(1)}, c_{m,l,2}^{(1)},\ldots, c_{m,l,\log_2\mathcal M}^{(1)}),
		      (c_{m,l,1}^{(2)}, c_{m,l,2}^{(2)},\ldots, c_{m,l,\log_2\mathcal M}^{(2)}), \ldots,\\
		     &(c_{m,l,1}^{(\widehat \tau)}, c_{m,l,2}^{(\widehat \tau)},\ldots, c_{m,l,\log_2\mathcal M}^{(\widehat \tau)}) 
		\},
\end{aligned}
\end{equation*}
the LLR of each bit, i.e., $c_{m,l,b}^{(j)}$ for $1 \le b \le \log_2 \mathcal M$, is calculated based on (\ref{e_Pro_fre}).
We have
\begin{equation} \label{e_Pro_bit}
	\begin{aligned}
	P \left( c_{m,l,b}^{(j)} \right) = 
	\sum_{ {\widehat \omega}_{m,l,i} \to c_{m,l,b}^{(j)}}	
	P \left( {\widehat \omega}_{m,l,i} \right),
	\end{aligned}
\end{equation}
indicating that the probability of the $b$th bit of ${\bf c}_{m,l}^{(j)}$
is decided by the mapping relationship among ${\bf c}_{m,l}^{(j)}$, ${x}_{m,l}^{(j)}$, and ${\widehat \omega}_{m,l}$, e.g., ${\bf c}_{m,l}^{(j)} \to {x}_{m,l}^{(j)} \to {\widehat \omega}_{m,l}$ and vice versa.

Then, we can obtain the initial LLR of $c_{m,l,b}^{(j)}$ as
\begin{equation} \label{e_LLR}
  \begin{aligned}
	\text{LLR} \left( c_{m,l,b}^{(j)} \right) = 
	{\rm Log} \left[ \frac{P\left(c_{m,l,b}^{(j)} = 0 \right)}
						  {P\left(c_{m,l,b}^{(j)} = 1 \right)} \right],
  \end{aligned}
\end{equation}
where $1 \le b \le \log_2 \mathcal M$, $1 \le l \le L$ and $1 \le m \le M$.
Thereafter, the LLRs of ${\bf C}^{(j)}$ are obtained.
By de-interleaving ${\bf C}^{(j)}$ column-by-column, we can achieve the soft LLRs of ${\bf v}^{(j)}$ that are used for further decoding.

\subsubsection{Improved MPA}

Assume the parity-check matrix of the first encoder is ${\bf H}_1$.
Since the second encoder utilizes SPC, encoders one and two can be jointly decoding.

Let the Tanner graph of the two encoders be $\mathcal G_1$ and $\mathcal G_2$, and the two graphs are connected by some of the variable nodes (VNs) of $\mathcal G_1$ and the check nodes (CNs) of $\mathcal G_2$. Thus, the VNs update and CNs update should take into consideration the joint Tanner graph $\mathcal G$, as shown in Fig. \ref{Fig4}. The proposed improved decoding algorithm is based on the joint Tanner graph $\mathcal G$, and the major alterations are the VNs of $\mathcal G_1$ update and CNs of $\mathcal G_2$ update.

\begin{itemize}
  \item
  VNs of $\mathcal G_1$ update. Since some of the VNs of $\mathcal G_1$ are connected by both the CNs of $\mathcal G_1$ and CNs of $\mathcal G_2$, the update LLRs of these VNs should taken into consideration both CNs set.
  \item
  CNs of $\mathcal G_2$ update. For the CNs of $\mathcal G_2$, it connects both its own VNs and some VNs of $\mathcal G_1$, thus, the update LLRs of the CNs of $\mathcal G_2$ are determined by the two VNs sets.
\end{itemize}

In the algorithm, there are some definitions of MPA.
The VN and CN nodes of $\mathcal G_1$ are defined by $\alpha_1$ and $\beta_1$;
similarly, the VN and CN nodes of $\mathcal G_2$ are set to be $\alpha_2$ and $\beta_2$.
Let $N(\alpha_1)$, $N(\alpha_2), N(\beta_1)$ and $N(\beta_2)$ be the sets that are respectively connected with VN node $\alpha_1$, VN node $\alpha_2$, CN node $\beta_1$ and CN node $\beta_2$.
Note that, there is no edge between $\alpha_2$ and $\beta_1$, because of the SPC structure.

In summary, the receiver should detect the number of arrival users, 
separate the superimposed signals, and recover the transmit signals of all the users.
The first step is to do SoF-AUD, followed by the MUD and improved MPA steps.
It is noted that, the initial soft LLR information of MPA is decided by the MUD step. 
In other words, there is a close relationship between MUD and MPA, sometimes, the MUD and MPA can be jointly considered. 
The entire detection algorithms are summarized in Algorithm 1.

\begin{tiny}
  \begin{algorithm} [t]
      \caption{Detection algorithms of the GFMA receiver.} \label{AlgorithmOne}
          \KwIn { ${\bf y}_m$ for $1 \le m \le M$, $N_{max}$. }
          \KwOut { ${\bf u}^{(j)}$ for $1 \le j \le J$. }
          \textbf{Step 1: SoF-AUD}:
          Obtain $\widehat \tau$ and $\widehat \mu$ by (\ref{e_L1andL2}); \\
          \textbf{Step 2: MUD} \\
          \If{without encoder ${\bf G}_1$}
          {
              Obtain ${\widehat \omega}_{m,l,1}$ by (\ref{e_MUD_hard}) when $\mathcal{M}_1 = 1$;
              and ${\widehat \omega}_{m,l}$ by (\ref{e_MUD_FSK_hard}) when $\mathcal{M}_1 > 1$. \\
              \textbf{Output}: Hard-decoding for $\mathbf{c}_m^{(j)}$, then obtain ${\bf u}^{(j)}$; \\
          }
          Obtain initial LLRs of $\alpha_1$ and $\alpha_2$ by (\ref{e_LLR}); \\
          \textbf{Step 3: Improved MPA} \\
          Set
          $\rm{LLR}_{\alpha_1 \rightarrow \beta_1 ({\rm or} \beta_2)}  = \rm{LLR}_{\alpha_1}$;
          $\rm{LLR}_{\alpha_2 \rightarrow \beta_2} = \rm{LLR}_{\alpha_2}$; Set $N_{sim} = 0$; \\
          \While{$N_{sim} < N_{max}$}
          {
              \textbf{Update CNs:} \\
  
              $\text{LLR}_{\beta_1 \rightarrow \alpha_1} = 2\tanh^{-1} \left[ \prod_{\alpha_1^{\prime} \in
              N(\beta_1)-\alpha_1} \tanh\left( \frac{1}{2} \text{LLR}_{\alpha_1^{\prime} \rightarrow \beta_1} \right) \right]$;
  
              $\text{LLR}_{\beta_2 \rightarrow \alpha_1} = 2\tanh^{-1} \left[ \prod_{\alpha_1^{\prime} \in
              N(\beta_2)-\alpha_1} \tanh\left( \frac{1}{2} \text{LLR}_{\alpha_1^{\prime} \rightarrow \beta_2} \right) \right]$;
  
              $\text{LLR}_{\beta_2 \rightarrow \alpha_2} = 2\tanh^{-1} \left[ \prod_{\alpha_2^{\prime} \in
              N(\beta_2)-\alpha_2} \tanh\left( \frac{1}{2} \text{LLR}_{\alpha_2^{\prime} \rightarrow \beta_2} \right) \right]$;

              \textbf{Update VNs:} \\
              ${\rm LLR}_{\alpha_1 \rightarrow \beta_1} = {\rm LLR}_{\alpha_1} + \sum_{\beta_1^{\prime}\in
              N(\alpha_1)-\beta_1} L_{\beta_1^{\prime} \rightarrow \alpha_1}$;
  
              ${\rm LLR}_{\alpha_1 \rightarrow \beta_2} = {\rm LLR}_{\alpha_1} + \sum_{\beta_2^{\prime}\in
              N(\alpha_1)-\beta_2} L_{\beta_2^{\prime} \rightarrow \alpha_1}$;
              ${\rm LLR}_{\alpha_2 \rightarrow \beta_2} = {\rm LLR}_{\alpha_2}$;
  
			  \textbf{Update LLR:} \\
			  ${\rm LLR_{\alpha_1}} = {\rm LLR_{\alpha_1}} + \sum_{\beta \in N(\alpha_1)} {\rm LLR}_{\beta_1 \rightarrow \alpha_1} + \sum_{\beta \in N(\alpha_2)} {\rm LLR}_{\beta_2 \rightarrow \alpha_1}$; \\

              Set $v_{n_1}^{(j)} = 1$ if $\text{LLR}_{\alpha_1} > 0$, otherwise, $v_{n_1}^{(j)} = 0$; \\
              \If{$N_{sim} \ge N_{\max} \: \| \: {\bf v}^{(j)} \cdot \mathbf{H}_1 = \mathbf{0}$}
              {
              \textbf{Output}: {${\bf v}^{(j)} \to {\bf u}^{(j)}$;} \\
              }
          }
      \end{algorithm}
\end{tiny}

\section{Theoretical Analysis}
In this section, we deduce theoretical AUER and Shannon limits for the proposed system.
The AUER is utilized to measure the performance of AUD.
Moreover, 
Shannon limits of the proposed system are derived, instead of the theoretical BER.

\subsection{AUER}
For a random access system, define active user error rate (AUER) by
\begin{equation}
P_{e, AU} = {\rm Pr}[\tau \neq {\widehat \tau}],
\end{equation}
where $\tau$ and $\widehat \tau$ respectively stand for the factual and detected numbers of arrival users.

In fact, the AUER of our proposed system is mainly determined by two aspects, 
one is the detection error caused by noise, 
and the other is the error floor caused by insufficient information. 
Our proposed detection algorithm is based on the sum power $\gamma_{sum}$, which is affected by several parameters that are the sum-pattern set of $\tau$ users $\Omega^{\tau}$, the number of rows $M$, 
and multi-dimensional order $\mathcal{M}_1$. 
Thus, the AUER can be derived as
\begin{equation}
P_{e,AU} = \sum_{\tau = 1}^{T} P_{\tau} \sum_{1 \le \tau' \le T, \tau' \neq \tau} \int_{D_{\tau'}} p(\gamma_{sum}|\tau) {\rm d} \gamma_{sum},
\end{equation}
where $P_{\tau}$ is {\it a priori} probability of the number of arrival users, 
$p(\gamma_{sum}|\tau)$ is the conditional probability density function (PDF) of $\gamma_{sum}$ given by the number of arrival users $\tau$ for $\tau = 1, 2, \ldots, T$, and $D_{\tau'}$ is the decision region of $\tau'$.

Thereafter, the key issue is to derive the PDF of $p(\gamma_{sum}|\tau)$.
Regarding as the sum-form of $\gamma_{sum}$ as shown in (\ref{e_gamma}), $p(\gamma_{sum}|\tau)$ is determined by a sequence of noncentral Chi-Square random variables, defined by
\begin{equation}  \label{gammma_PDF}
p(\gamma_{sum}|\tau) = \sum_{s_a \in \mathcal S} P(s_a) \cdot f_{\chi^2}(s_a),
\end{equation}
where $f_{\chi^2}(s_a)$ is the PDF of a noncentral Chi-Square random variable $\gamma_s$, 
where $\gamma_s = M \cdot \gamma_{sum} $, given by 
\begin{equation} \label{chi_PDF}
f_{\chi^2}(s_a)= \frac{1}{N_0} \left( \frac{\gamma_{s}}{s_a^2} \right)^{\frac{N_f-2}{4}} 
      {\rm e}^{-\frac{s_a^2+\gamma_{s}}{N_0}} \cdot 
      {\rm I}_{\frac{N_f}{2}-1}\left( \frac{s_a}{N_0 / 2}\sqrt{\gamma_{s}} \right), 
\end{equation}
where ${\rm I}_{\alpha}(r) = \sum_{t=0}^{\infty} \frac{ (r/2)^{\alpha+2t}  }{t! \Gamma(\alpha + t + 1)}$ 
and $\Gamma(r) = \int_{0}^{\infty} t^{r-1} e^{-t} dt$.
$\gamma_s$ is consisted of $N_f$ independent Gaussian variables with comment variance $N_0/2$ and different means denoted by $s_{n_f}$.
Since $r_{m,l,i}$ is a complex number, the degree of freedom is $N_f = 2L \cdot M$.

In (\ref{gammma_PDF}), $s_a$ is defined by $s_a =  \sqrt{\sum_{n_f=1}^{N_f} s_{n_f}^2}$, 
and all the values of $s_a$ are from the set $\mathcal S$, i.e., $s_a \in \mathcal S$. 
The probability of $s_a$ in set $\mathcal S$ is defined by $P(s_a)$, thus, $\sum_{s_a \in \mathcal S} P(s_a) = 1$.

Then, analyzing $p(\gamma_{sum}|\tau)$ is equivalent to finding the values of $s_a$ and $P(s_a)$, which can be calculated by the sum-pattern set $\Omega^{\tau}$.
Let us review the expression of $\Omega^{\tau}$, which is defined by $\Omega^{\tau} = \{\Omega_1^{\tau}, \Omega_2^{\tau}, \ldots, \Omega_l^{\tau},\ldots, \Omega_L^{\tau} \}$ with $\Omega_l^{\tau} = \omega_l^{(\tau,1)} \cup \ldots \cup \omega_l^{(\tau,\mu)} \cup \ldots \omega_l^{(\tau, C_T^{\tau})}$.

Divide the complex number set $\omega_l^{(\tau,\mu)}$ into two sets, the real number set $\omega_{l,re}^{(\tau,\mu)}$ and the image number set $\omega_{l,im}^{(\tau,\mu)}$. 
Evidently, $s_{n_f}$ belongs to the set $\omega_{l,re}^{(\tau,\mu)} \cup \omega_{l,im}^{(\tau,\mu)}$, 
i.e., $s_{n_f} \in \omega_{l,re}^{(\tau,\mu)} \cup \omega_{l,im}^{(\tau,\mu)}$, 
varying with the varied $\tau$ and $\mu$, so to the variable $s_a$. 
For example, if $\tau = 1$, it is known that 
$s_a = \sqrt{\sum_{n_f=1}^{N_f} s_{n_f}^2} = \sqrt{M\sum_{l=1}^{L} a_{l}^2} = \sqrt{ ML\cdot P_{avg}}$.

In general, $s_a$ and $P(s_a)$ are both affected by the ratio of the number of elements in $\omega_l^{(\tau,\mu)}$ to the total number of elements in $\Omega_l^{\tau}$. Therefore, different UDAS sets result in different $s_a$ and $P(s_a)$, thus obtaining different $p(\gamma_{sum}|\tau)$.

To explain the analysis processing of $p(\gamma_{sum}|\tau)$, we present an example. 
If we exploit the UDAS set given by Example 2, as shown in (5), we have
\begin{small}
\begin{equation} \label{tau_2_PDF}
  \begin{aligned}
    p(\gamma_{sum}|\tau=1) &= f_{\chi^2}(\sqrt{ ML\cdot P_{avg}}),\\
    p(\gamma_{sum}|\tau=2) &= \frac{2}{3} f_{\chi^2}(\sqrt{20M}) + \frac{1}{3} \sum_{k = 0}^{LM} \left[ p(k, \frac{N_f}{2}) f_{\chi^2}(\sqrt{9k + (LM-k)}) \right],\\
    p(\gamma_{sum}|\tau=3) &= \sum_{k = 0}^{LM} \left[ p(k, \frac{N_f}{2}) f_{\chi^2}(\sqrt{9k + (LM-k) + 10M}) \right],\\
    p(\gamma_{sum}|\tau=4) &= \sum_{k = 0}^{N_f} \left[ p(k, N_f) f_{\chi^2}(\sqrt{9k + (N_f-k)}) \right],\\
  \end{aligned}
\end{equation}
\end{small}
where $p(k, N_f) = C_{N_f}^{k} \cdot (\frac{1}{2})^{N_f}$ is the probability of a binomial distribution that 
selects $k$ from $N_f$. 
The case of $\tau=1$ is accord with the aforementioned discussion.

When $\tau = 2$, it is known that $\mu$ belongs to $[1, C_T^{\tau}]$, and $\Psi^{(2, \mu)}$ is given in Example 4. Specifically, they can be divided into the following two cases.
\begin{enumerate}
  \item
  Case 1: $\mu=1, 3, 4$ or $6$.
  At the moment, the sum-patterns of the two users include both real and image numbers.
  Take $\mu = 1$ as an example, we can get 
  $\omega_1^{(2,1)} = \{1+2i, 1-2i, -1+2i, -1-2i \}$, 
  $\omega_2^{(2,1)} = \{1+1i, 1-1i, -1+1i, -1-1i \}$, 
  $\omega_3^{(2,1)} = \{2+1i, 2-1i, -2+1i, -2-1i \}$ and $\omega_4^{(2,1)} = \{2+2i, 2-2i, -2+2i, -2-2i \}$.
  Therefore, the power of the sum-pattern always keeps as a constant. 
  For example, the power values of the four symbols ($l=1,2,3,4$) are respectively 5, 2, 5, and 8.
  Then, $s_a$ is calculated as $s_a = \sqrt{M\cdot (5 + 2 + 5 + 8)} = \sqrt{20M}$. 
  The cases of $\mu = 3,4$ and $6$ are exactly the same as the case of $\mu=1$.
  \item
  Case 2: $\mu=2$ or $5$.
  At the moment, the sum-patterns of the two users are all real numbers (or image numbers).
  Obviously, we can get 
  $\Omega^{(2,2)} = \{ +3, +1, -1, -3\}$ and 
  $\Omega^{(2,5)} = \{ +3i, +1i, -1i, -3i\}$. 
  Now, we take $\mu=2$ as an example to analyze.
  Since the sum-patterns of $\mu=2$ are all real numbers, then $LM$ Gaussian variables of the totally $N_f$ degrees are distributed as ${\mathcal N}(0, N_0/2)$, and the others $LM$ variables are in probability distributed as ${\mathcal N}(\pm 3, N_0/2)$ or ${\mathcal N}(\pm 1, N_0/2)$.
  Assume $k$ Gaussian variables are distributed as ${\mathcal N}(\pm 3, N_0/2)$, 
  and the rest $LM - k$ Gaussian variables are with distribution as ${\mathcal N}(\pm 1, N_0/2)$, 
  where $0 \le k \le LM$ with probability $C_{LM}^{k} (\frac{1}{2})^{LM}$. 
  Thereafter, $s_a$ is calculated as 
  $s_a = \sqrt{k \cdot 9 + (LM-k)\cdot 1 + LM \cdot 0} = \sqrt{9k + (LM-k)}$.
\end{enumerate}
Consider the probabilities of cases 1 and 2 are respectively $2/3$ and $1/3$, we can derive the final PDF of $p(\gamma_{sum}|\tau=2)$ as shown in (\ref{tau_2_PDF}). 
When $\tau > 2$, the analysis processing is the same as the case of $\tau =2$, which will not be repeated described.

It is remarkable that the detection error probability of the UDAS set $\Psi^{(\tau,\mu)}$ can be ignored compared to the AUER.

\subsection{Shannon limit of the proposed system in an adder multiple-access channel}
The computation of Shannon limit involves two steps, 
calculating the channel capacity and solving the lowest $E_b/N_0$, 
where $E_b$ represents the bit energy of each user.

Assume $R_j$ is the data rate of the $j$th user, and $P_j$ is the transmit power of the $j$th user.
Then, the capacity of a MAC is given by \cite{Book_Info}
\begin{equation} \label{e_MAC}
	\begin{aligned}
		\sum_{j=1}^{J} R_j \le \log_2 \left(1 + \frac{P_1+P_2+\ldots+P_J}{N_0} \right).
	\end{aligned}
\end{equation}
If all the users hold the same data rate, i.e.,  $R_1=R_2=\ldots=R_J=R_c$, 
the last inequality dominates the others. 
It is remarkable that the Shannon limit of the proposed system is deduced based on this assumption.
Besides, due to the cyclic UDAS set, it is able to known $P_1 = P_2 =\ldots=P_J = P_{avg}$.
Thus, the relationship between $P_{avg}$ and $E_b$ is 
$P_{avg} = \frac{R_c \cdot \log_2(\mathcal M) \cdot E_b}{T_s}$, where $T_s$ is the symbol duration.

In the following discussion, assume the number of arrival users $J = \tau$ and the selected UDAS set $\Psi^{(\tau, \mu)}$, i.e., $\Psi^{(\tau, \mu)} = \{ {\bf e}_{t_1}, {\bf e}_{t_2}, \ldots, {\bf e}_{t_\nu}, \ldots, {\bf e}_{t_\tau} \}$ with $t_\nu \in \{1, 2, \ldots, T\}$, have been available.
Denote $\{ {\vec \chi}_{l,m_s}^{(j)}, 1 \le m_s \le {\mathcal M} \}$ by the transmit constellation set of the $l$th symbol of the $j$th user, including ${\mathcal M}$ possible ${\mathcal M_1}$-dimension transmit signals, expressed as
${\vec \chi}_{l,m_s}^{(j)} = \left\{ {\chi}_{l,m_s,1}^{(j)}, {\chi}_{l,m_s,2}^{(j)}, \cdots,
{\chi}_{l,m_s,i}^{(j)}, \cdots, {\chi}_{l,m_s,\mathcal M_1}^{(j)} \right\}$,
where ${\chi}_{l,m_s,i}^{(j)} \in \{-e_{t_j,l}, 0, e_{t_j,l} \}$, 
with the assumption that the $j$th user selects the UDAS ${\bf e}_{t_j}$ in $\Psi^{(\tau, \mu)}$.
Define ${\rm Pr} \left( {\vec \chi}_{l,m_s}^{(j)} \right)$ by the {\textit {prior}} probability of ${\vec \chi}_{l,m_s}^{(j)}$.

When $J$ users' signals, i.e., ${\vec \chi}_{l,m_s}^{(j)}$, are transmitted through a MAC, 
the PDF of the $l$th symbol of the received signal ${\bf y}_{m,l}$ is 
\begin{small}
\begin{equation}\label{e_PDF_y}
  \begin{aligned}
  p\left( {\bf y}_{m,l}|{\vec \chi}_{l,m_s}^{(1)}, {\vec \chi}_{l,m_s}^{(2)}, \ldots, {\vec \chi}_{l,m_s}^{(J)} \right)
    &= \mathop {\prod \limits_{m_s = 1}^{\mathcal M} \ldots \prod \limits_{m_s = 1}^{\mathcal M}}
       \limits_{\mathcal M - {\rm{fold}}}
       {\frac{1}{{\sqrt {\pi {N_0}} }}
       \exp \left\{ { - \frac{{{{\left( {\bf y}_{m,l} - \sum_{j=1}^{J} {\vec \chi}_{l,m_s}^{(j)} \right)}^2}}}
       {{{N_0}}}}\right\}}.
  \end{aligned}
\end{equation}
\end{small}
Note that the row index has no effect to the PDF.

\begin{table} [t]
  \begin{center}
    \caption{Shannon limits of the proposed system in an adder multiple-access channel, where $\mathcal M_1 = 1$, $J=2$, $L=4$, and $\Psi^{(2,1)} = \{(1, 1i, 2, 2i), (2i, 1, 1i, 2)\}$. The unit of $(\frac{E_b}{N_0})_{\min}$ is dB.}
    \begin{tabular}{|c c|c c|c c|c c|c c|c c|}
    \hline
    $R_c$ & $(\frac{E_b}{N_0})_{\min}$ & $R_c$ & $(\frac{E_b}{N_0})_{\min}$ &
    $R_c$ & $(\frac{E_b}{N_0})_{\min}$ & $R_c$ & $(\frac{E_b}{N_0})_{\min}$ &
    $R_c$ & $(\frac{E_b}{N_0})_{\min}$ & $R_c$ & $(\frac{E_b}{N_0})_{\min}$ \\
    \hline
  0.001 & -1.5858 & 0.151 & -0.9594 & 0.435 & 0.4667 & 0.646 & 2.0022 & 0.794 & 3.6701 & 0.895 & 5.4437 \\
  0.006 & -1.5682 & 0.175 & -0.8524 & 0.459 & 0.6154 & 0.663 & 2.1577 & 0.804 & 3.8150 & 0.901 & 5.5703 \\
  0.013 & -1.5392 & 0.226 & -0.6215 & 0.483 & 0.7657 & 0.694 & 2.4673 & 0.814 & 3.9586 & 0.921 & 6.0639 \\
  0.022 & -1.4992 & 0.250 & -0.4985 & 0.506 & 0.9176 & 0.709 & 2.6212 & 0.823 & 4.1007 & 0.934 & 6.4212 \\
  0.054 & -1.3716 & 0.305 & -0.2394 & 0.550 & 1.2248 & 0.736 & 2.9263 & 0.840 & 4.3807 & 0.951 & 6.9939 \\
  0.066 & -1.3185 & 0.330 & -0.1042 & 0.571 & 1.3797 & 0.749 & 3.0774 & 0.856 & 4.6550 & 0.962 & 7.4329 \\
  0.106 & -1.1538 & 0.384 & 0.1759  & 0.610 & 1.6907 & 0.773 & 3.3762 & 0.877 & 5.0556 & 0.981 & 8.4635 \\
  0.128 & -1.0600 & 0.416 & 0.3566  & 0.628 & 1.8465 & 0.784 & 3.5238 & 0.883 & 5.1864 & 0.991 & 9.3165 \\
    \hline
    \end{tabular}
    \vspace{-25pt}
  \end{center}
\end{table}

Consider the transmit mode of each user, the channel capacity of the $l$th symbol in the adder MAC can be calculated as
\begin{equation}\label{e_capacity_MAC}
  \begin{aligned}
  C_l &= \max {\text I} \left( \mathcal Y ; \mathcal X_1, \mathcal X_2, \ldots, \mathcal X_J \right) 
      = \mathop {\max } 
    \mathop {\int\limits_{ - \infty }^\infty   \cdots  \int\limits_{-\infty }^\infty}
    \limits_{\mathcal M - {\rm{fold}}}
    p\left( {\bf y}_{m,l}|{\vec \chi}_{l,m_s}^{(1)}, {\vec \chi}_{l,m_s}^{(2)}, \ldots,
          {\vec \chi}_{l,m_s}^{(J)} \right) \cdot \\
    &{\rm Pr} \left( {\vec \chi}_{l,m_s}^{(1)} \right) \cdots
     {\rm Pr} \left( {\vec \chi}_{l,m_s}^{(J)} \right) \cdot 
     {\log _2}
     \left[ \frac{ p\left( {\bf y}_{m,l}|{\vec \chi}_{l,m_s}^{(1)}, 
     	{\vec \chi}_{l,m_s}^{(2)}, \ldots, 
     	{\vec \chi}_{l,m_s}^{(J)} \right) }
     	{p\left( {\bf y}_{m,l} \right)} \right] {\rm d}{\bf y}_{m,l}. \\
  \end{aligned}
\end{equation}
where $\mathcal Y$ is the the received signal's variable, 
$\mathcal X_1, \mathcal X_2, \ldots, $ and $\mathcal X_J$ are the transmit signals' variables of $J$ users.

Since each user exploits an $L$-length UDAS, we define ergodic capacity as
\begin{equation}
{\overline C} = {\rm E}[C_l] = \frac{1}{L} \sum_{l=1}^{L} C_l.
\end{equation}

Regarding as (\ref{e_MAC}), it is known that
\begin{equation}
R_c \cdot {\log_2} {\mathcal M} \cdot J \le {\overline C} = f\left( {{{{E_b}} \mathord{\left/ {\vphantom {{{E_b}} {{N_0}}}} \right. \kern-\nulldelimiterspace} {{N_0}}}} \right).
\end{equation}

\begin{table} [t]
  \begin{center}
      \caption{Shannon limits of the proposed system in an adder multiple-access channel, where $\mathcal M_1 = 1$, $J=3$, $L=4$, $\Psi^{(3,1)} = \{(1, 1i, 2, 2i), (2i, 1, 1i, 2), (2, 2i, 1, 1i)\}$. The unit of $(\frac{E_b}{N_0})_{\min}$ is dB.}
      \begin{tabular}{|c c|c c|c c|c c|c c|c c|}
        \hline
        $R_c$ & $(\frac{E_b}{N_0})_{\min}$ & $R_c$ & $(\frac{E_b}{N_0})_{\min}$ &
        $R_c$ & $(\frac{E_b}{N_0})_{\min}$ & $R_c$ & $(\frac{E_b}{N_0})_{\min}$ &
        $R_c$ & $(\frac{E_b}{N_0})_{\min}$ & $R_c$ & $(\frac{E_b}{N_0})_{\min}$ \\
        \hline
  0.001 & -1.5840 & 0.214 & -0.3740 & 0.499 & 1.6466 & 0.690 & 3.4298 & 0.813 & 4.8805 & 0.907 & 6.4123 \\
  0.006 & -1.5609 & 0.237 & -0.2279 & 0.518 & 1.8023 & 0.703 & 3.5688 & 0.822 & 5.0045 & 0.912 & 6.5236 \\
  0.013 & -1.5229 & 0.261 & -0.0785 & 0.536 & 1.9571 & 0.716 & 3.7063 & 0.830 & 5.1274 & 0.922 & 6.7438 \\
  0.022 & -1.4706 & 0.284 & 0.0736  & 0.554 & 2.1106 & 0.728 & 3.8423 & 0.847 & 5.3699 & 0.931 & 6.9608 \\
  0.049 & -1.3268 & 0.330 & 0.3839  & 0.587 & 2.4139 & 0.751 & 4.1100 & 0.862 & 5.6081 & 0.953 & 7.5932 \\
  0.065 & -1.2371 & 0.353 & 0.5411  & 0.604 & 2.5635 & 0.763 & 4.2417 & 0.869 & 5.7258 & 0.956 & 7.6959 \\
  0.103 & -1.0277 & 0.397 & 0.8575  & 0.634 & 2.8583 & 0.784 & 4.5011 & 0.883 & 5.9582 & 0.971 & 8.2967 \\
  0.167 & -0.6532 & 0.439 & 1.1745  & 0.663 & 3.1471 & 0.803 & 4.7552 & 0.895 & 6.1870 & 0.990 & 9.5969 \\
    \hline
    \end{tabular}
    \vspace{-20pt}
\end{center}
\end{table}

\begin{table} [t]
  \begin{center}
    \caption{Shannon limits of the proposed system in an adder multiple-access channel, where $\mathcal M_1 = 1$, $J=2$, $L=6$, and $\Psi^{(2,1)} = \{(1, 1i, 2, 2i, 4, 4i), (4i, 1, 1i, 2, 2i, 4)\}$.
    The unit of $(\frac{E_b}{N_0})_{\min}$ is dB.}
    \vspace{-5pt}
    \begin{tabular}{|c c|c c|c c|c c|c c|c c|}
\hline
$R_c$ & $(\frac{E_b}{N_0})_{\min}$ & $R_c$ & $(\frac{E_b}{N_0})_{\min}$ &
$R_c$ & $(\frac{E_b}{N_0})_{\min}$ & $R_c$ & $(\frac{E_b}{N_0})_{\min}$ &
$R_c$ & $(\frac{E_b}{N_0})_{\min}$ & $R_c$ & $(\frac{E_b}{N_0})_{\min}$ \\
\hline
0.004&  -1.5693 &  0.275& 0.0848  & 0.574& 2.9037 & 0.726& 5.1081 & 0.914& 9.1845 & 0.833& 7.0888 \\
0.016&  -1.5033 &  0.312& 0.3579  & 0.595& 3.1718 & 0.739& 5.3254 & 0.918& 9.3288 & 0.841& 7.2666 \\
0.035&  -1.3973 &  0.348& 0.6377  & 0.615& 3.4343 & 0.752& 5.5379 & 0.922& 9.4709 & 0.848& 7.4413 \\
0.090& -1.0846  &  0.415& 1.2092  & 0.651& 3.9415 & 0.775& 5.9492 & 0.941& 10.1529& 0.863& 7.7819 \\
0.124&  -0.8872 &  0.446& 1.4972  & 0.668& 4.1862 & 0.786& 6.1485 & 0.950& 10.5407& 0.876& 8.1111 \\
0.198&  -0.4309 &  0.502& 2.0699  & 0.698& 4.6582 & 0.806& 6.5354 & 0.971& 11.6227& 0.894& 8.5855 \\
0.236&  -0.1790 &  0.529& 2.3521  & 0.713& 4.8858 & 0.815& 6.7233 & 0.981& 12.2853& 0.904& 8.8895 \\
\hline
\end{tabular}
\end{center}
\vspace{-25pt}
\end{table}


When $R_c \cdot {\log_2} {\mathcal M} \cdot J = \overline C$, the utilization of communication resources is maximized. 
The Shannon limit is defined as the minimum $E_b/N_0$ that can realize reliable transmission with $R_c$, which is denoted by ${\left( {{{{E_b}} \mathord{\left/ {\vphantom {{{E_b}} {{N_0}}}} \right. \kern-\nulldelimiterspace} {{N_0}}}} \right)_{\min }}$. Thus,
\begin{equation} \label{e_min_Eb}
{\left( {{{{E_b}} \mathord{\left/ {\vphantom {{{E_b}} {{N_0}}}} \right. \kern-\nulldelimiterspace} {{N_0}}}} \right)_{\min }} = {f^{ - 1}}\left( R_c \cdot \log_2 \mathcal M \cdot J \right).
\end{equation}

However, the integral in (\ref{e_capacity_MAC}) is extremely complex, especially the inverse function. 
Since it is one-to-one mapping between the code rate $R_c$ and $(E_b/N_0)_{\rm min}$, 
the corresponding code rate $R_c$ can be calculated for a given $(E_b/N_0)_{\rm min}$.
Then, we can obtain a sequence of data points $\left(R_c, (E_b/N_0)_{\rm min}\right)$.
In practice, the range of the integral and the accuracy of interpolation are depended on the required accuracy of the Shannon limit.


Table I, Table II and Table III show the Shannon limits of the proposed system in an adder multiple-access channel. When $R_c$ and $L$ are given, it is found that the required $\left({E_b}/{N_0} \right)_{\min}$ of the $J = 3$ case is larger than the $J=2$ case. Moreover, for a given number of arrival users $J$ and a given $R_c$, a larger $L$ indicates a larger $\left({E_b}/{N_0} \right)_{\min}$, because of the reduced minimum distance of the constellation at the receiver.

\section{Simulation Results}
In this section, we simulate and compare the performances of the proposed system.
The UDAS sets are generated based on the cyclic matrix mode, with generators ${\bf a} = (1, 1i, 2, 2i)$ of $L=4$, and ${\bf a} = (1, 1i, 2, 2i, 4, 4i)$ of $L=6$.
Moreover, the first encoder exploits a (3,6)-regular QC-LDPC (1016,508) with rate 0.5, and the second encoder is a SPC with rate $R_2 = 1 - \frac{1}{L \cdot \log_2{\mathcal M}}$. 
Thus, the total data rate of each user is equal to $R_c = 0.5 \times (1 - \frac{1}{L \cdot \log_2{\mathcal M}})$.
For example, when $L = 4$ and $\mathcal M = 2$, it is found that $R_c = 0.375$.

\begin{figure}
  \centering
    \includegraphics[width=8.0cm]{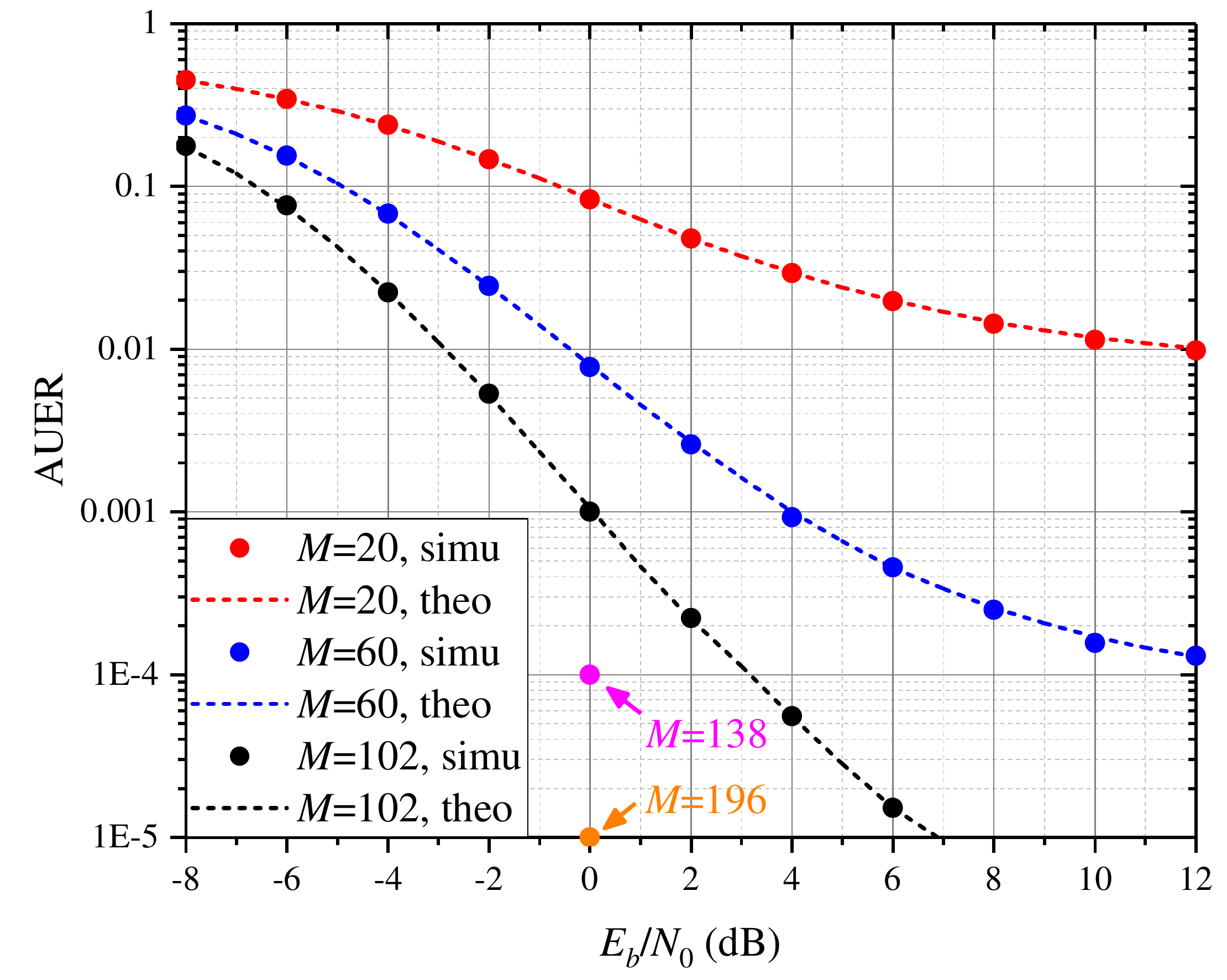}
    \caption{AUER of the proposed system with various parameters, where $\mathcal{M}_1 = 1$, $L = 4$, and $M = 20, 60, 102, 138$ and $196$.}\label{Fig5}
\end{figure}

First of all, we focus on the AUER performance, 
with the assumption that $P_{\tau}$ is uniform distribution in $[1, T]$.
To observe the AUER performance, we set $R_c = 1$ and omit the two encoders.
The AUER performance of the proposed system is shown in Fig. \ref{Fig5}.

It can be seen from Fig. \ref{Fig5} that the AUER decreases with the increased $E_b/N_0$.
When $M$ is small, e.g., $M=20$, there exists a significantly error floor, because of the lack of statistical information. 
With the increase of $M$, the AUER improves significantly, revealing that sufficient statistical information may reduce both the influence of noise and error floor.
Moreover, the simulated results are perfect accord with our deduced theoretical results.

When we set $E_b/N_0 = 0$ dB and AUER $\thickapprox 10^{-3}, 10^{-4}$ and $10^{-5}$, it is found that the minimum required numbers of $M$ are respectively 102, 138, and 196, whose corresponding transmit blocks are with length 408, 552 and 784.
Evidently, these block lengths satisfy the requirement of short packet communications. 
Therefore, we can maximum simultaneously support 4 users with a $10^{-5}$ detected error rate, 
at this moment, the required $E_b/N_0$ is only 0 dB. This result is appealing for a random access network.



\begin{figure}[t]
  \centering
  \subfigure{
  \includegraphics[width=3.2in]{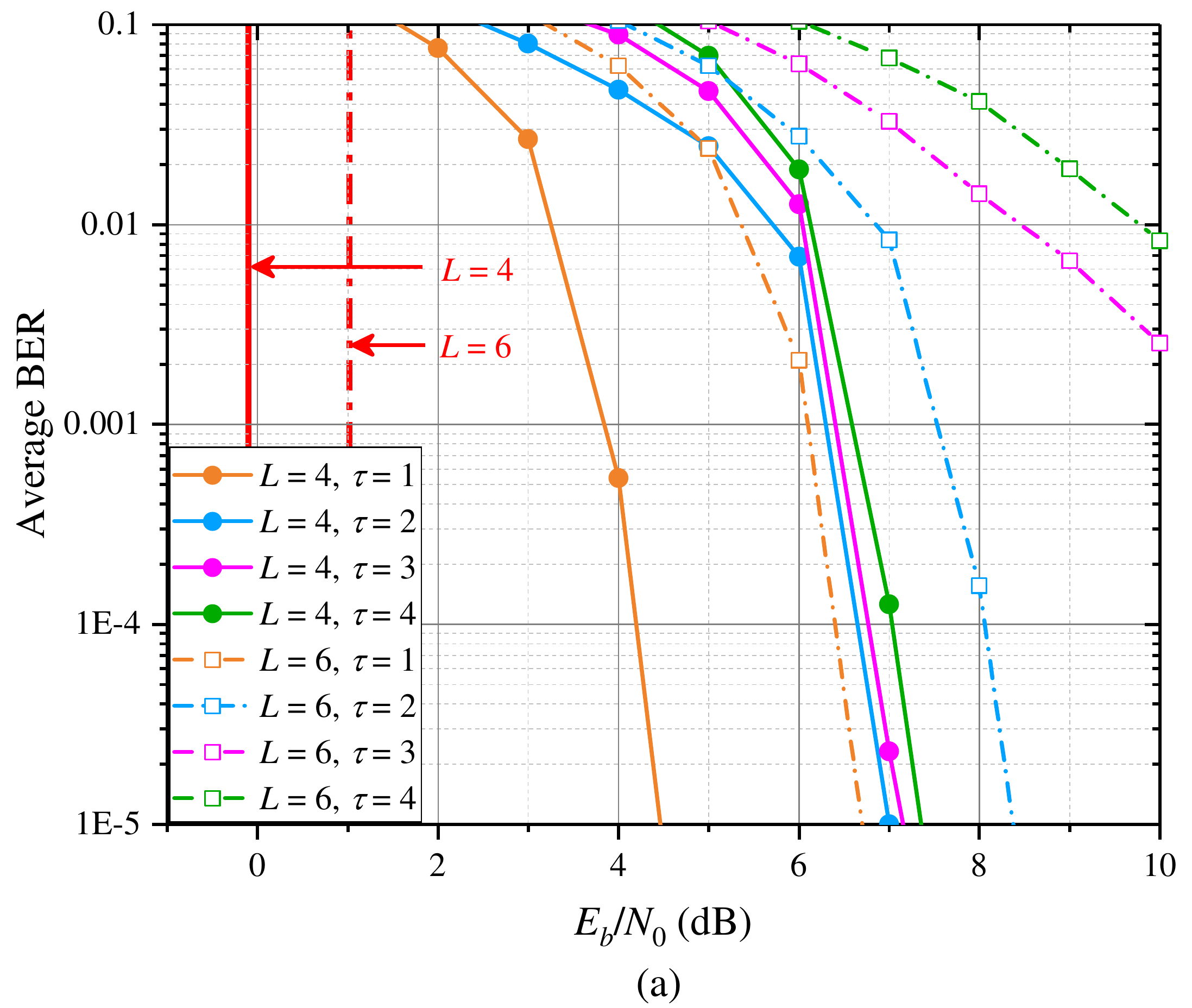}
  \label{6-a}
  }%
  \subfigure{
  \includegraphics[width=3.2in]{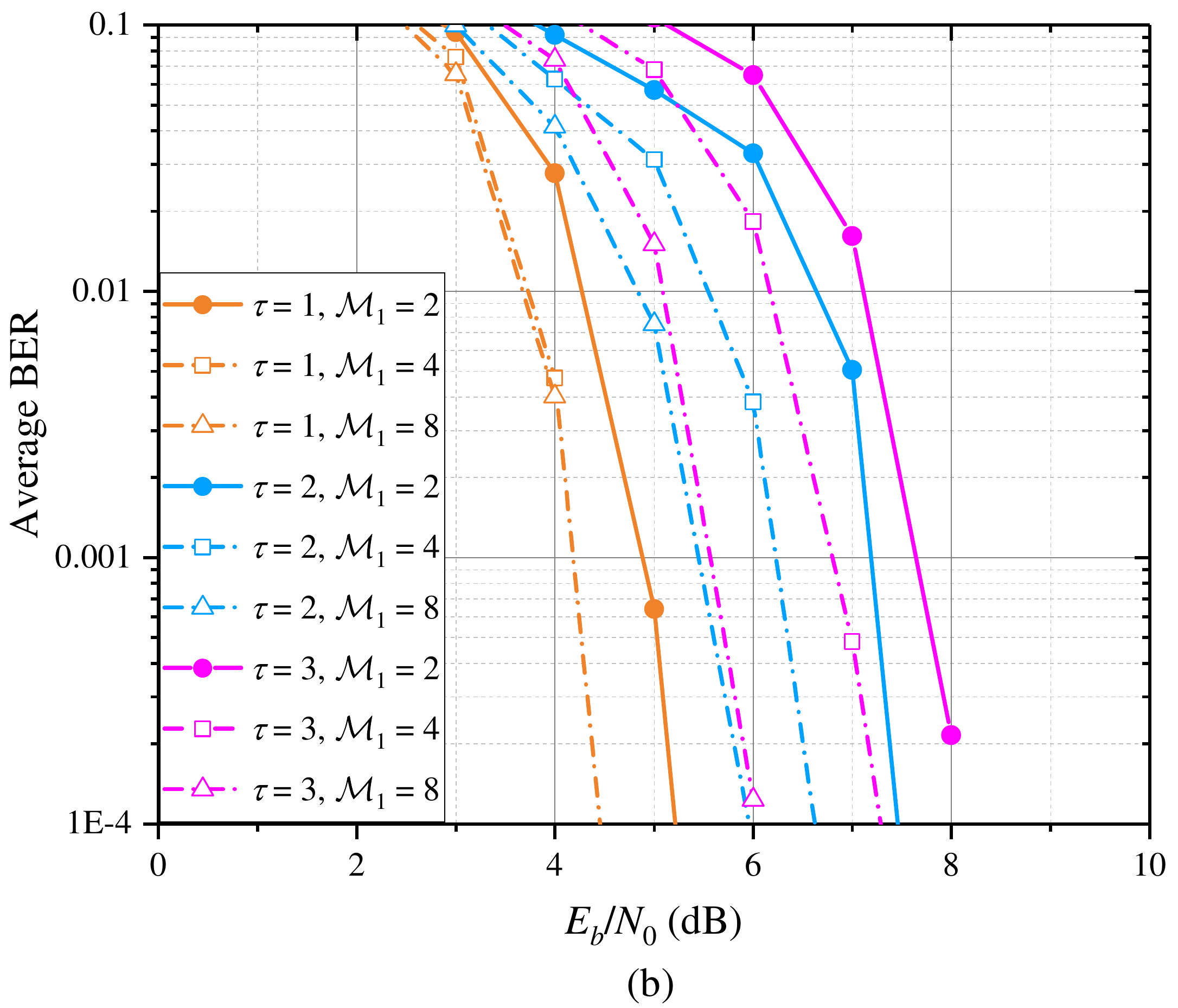}
  \label{6-b}
  }%
  \centering
  \caption{Average BER performances of the proposed system, where the maximum iteration number is 10. 
  (a) $M_1 = 1$ case, with $L=4, 6$, and $\tau = 1, 2, 3, 4$. 
  (b) $M_1 > 1$ case, with $L=4$, and $\tau = 1, 2, 3$.}
  \label{Fig6}
  \end{figure}

The BER performances of the proposed system are shown in Fig. \ref{Fig6}, where the block length of the transmit block is equal to or larger than 1016 that is the codeword length of the first encoder. Therefore, we assume that the number of arrival users and UDAS set have been perfectly detected.

Fig. \ref{6-a} shows the BER performance of the proposed scheme with parameter $\mathcal M_1=1$.
When $L=4$, it is found that BER of the case $\tau=1$ provides the best performance, following by the cases of $\tau=2$, $\tau=3$ and $\tau=4$. However, the BER differences among $\tau=2, 3, 4$ are small.
When BER is $10^{-4}$ and $L=4$, the BER of $\tau=3$ is about 6.8 dB away from the Shannon limit, and 6.7 dB away from the Shannon limit for the case of $\tau=2$.
Therefore, there exists an improvement space to further reduce the BER.
For a given $\tau$, e.g., $\tau=2$, the BER of the case of $L=4$ is much better than the case of $L=6$, since the received superimposed signal of the $L=4$ case has a relative large minimum Euclidean distance.
For example, when $\tau=2$ and BER is $10^{-4}$, the case of $L=4$ is about 1.5 dB better than that of the case of $L=6$. This result is accord with our deduced Shannon limits.


The BER performance of the proposed scheme with various dimensions $\mathcal M_1=2, 4, 8$ 
is shown in Fig. \ref{6-b}.
It can be seen that the BER performance is improved with the increased $\mathcal{M}_1$, 
due to the statistic feature and larger Euclidean distance.
For example, when $\tau=2$ and BER is $10^{-3}$, the case of $\mathcal{M}_1 = 4$ provides about 0.7 dB gain than the case of $\mathcal{M}_1 = 2$.

\begin{figure}[tbp]
  \centering
  \includegraphics[width=8.5cm]{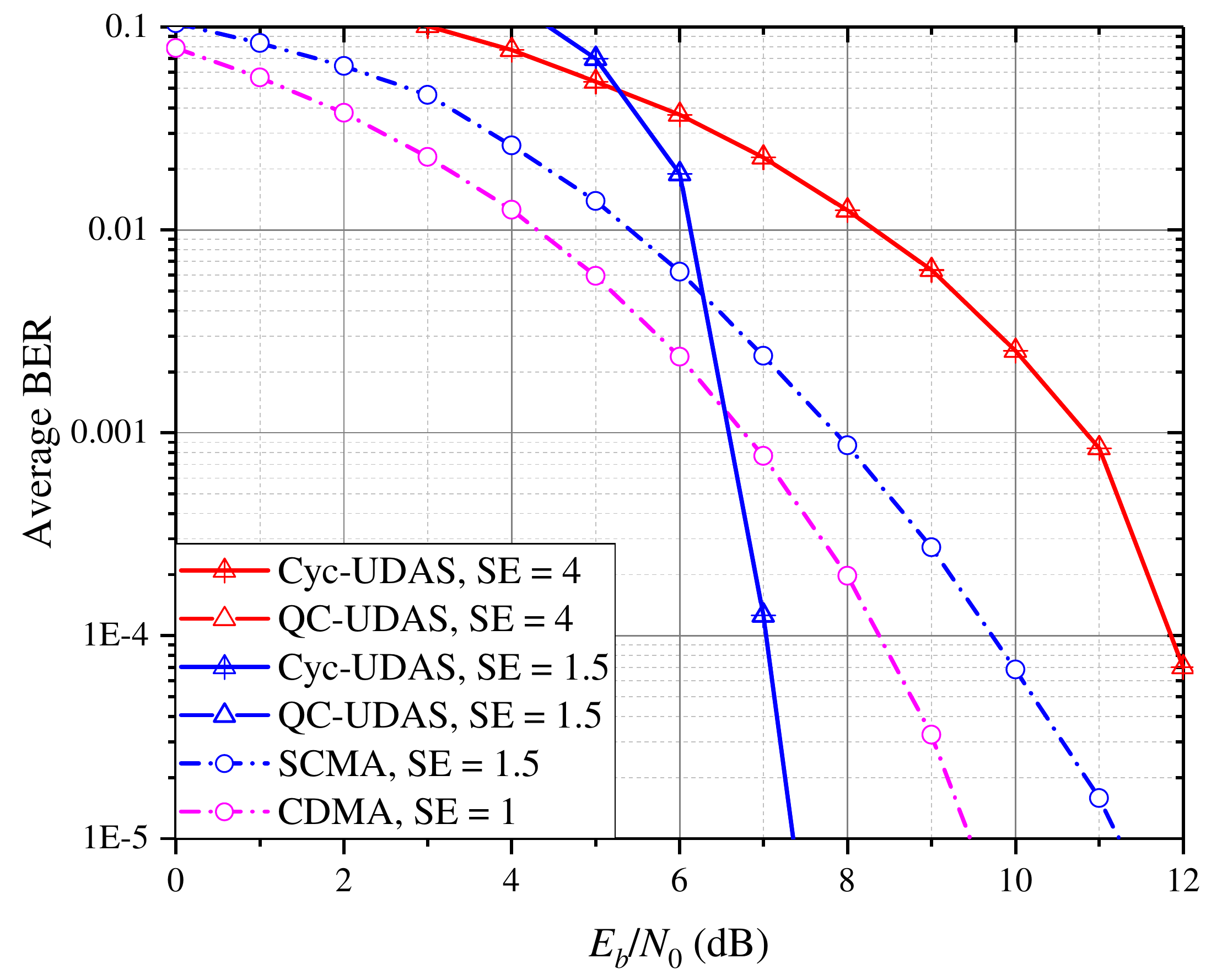}
  \caption{BER comparisons among different systems. The CDMA system utilizes Walsh sequences whose SF is 4, and SCMA system is operated based on the codebook given by \cite{GF_SCMA_Codebook1_2017}. Both the schemes one and two of our proposed systems have the same parameters, where $L=4$, $\mathcal M_1=1$, and $\tau = 4$. The scheme one is without any encoder; and scheme two is with two encoders. Both cyclic and quasi-cyclic structures are taken into consideration.}
  \label{Fig7}
\end{figure}

Fig. \ref{Fig7} shows the BER comparisons among different systems. 
We compare our proposed system with the classical CDMA system that exploits Walsh sequences whose SF is 4, 
and the SCMA system that is based on the codebook given by \cite{GF_SCMA_Codebook1_2017}.
Therefore, the SEs of the CDMA and SCMA systems are respectively 1 bit/resource and 1.5 bits/resource.
Our proposed system considers two schemes for comparisons.
The scheme one is corresponding to the parameters of $L=4$, $\mathcal M_1=1$, $\tau = 4$ without any encoders, indicating that the SE of case one is equal to 4 bits/resource.
On the other hand, the scheme two has the same parameters as that of the scheme one, and the only difference is caused by the inserted encoders, resulting in the SE of scheme two becomes as 1.5 bits/resources.

It is found that, the BER of our proposed scheme one is worse than those of CDMA and SCMA systems.
However, our proposed scheme one provides much higher SE.
Based on the inserted channel codes, our proposed scheme two can provide much better BER performances than those of CDMA and SCMA, when $E_b/N_0$ is larger than 6.6 dB.
At this time, our proposed scheme two has the same SE as that of SCMA system, verifying the validity of our proposed system.
Moreover, both the cyclic and quasi-cyclic UDAS modes provide the same BER performances, since they have the same UDAS set.

\section{Conclusion}
This paper introduces UDAS for grant-free multiple-access systems.
First of all, we present an UDAS-based MD-BICM transmitter for each user, which is a combination of two channel encoders, one interleaver and multi-dimensional modulation. The first encoder is used to improve the reliability of the system, and the second encoder is used for assisting AUD. 
Following, some definitions of UDAS are introduced in details. Refer to the constructions of QC-LDPC, two kinds of UDAS sets are presented, which are cyclic and quasi-cyclic matrix modes.
The cyclic/quasi-cyclic structure can help the receiver realize low-complexity AUD.
Thirdly, we present a SoF-AUD, and a joint MUD and improved MPA for the proposed system, where the Tanner graph of the decoder is the combination of two encoders $\mathcal G_1$ and $\mathcal G_2$.
Finally, both the theoretical AUER and Shannon limits are deduced in details.
We simulate and compare our proposed system with various parameters and systems.
When $E_b/N_0 = 0$ dB and the length of transmit block is larger than a given value, e.g., 784, the AUER of our proposed system can be an extremely low value $10^{-5}$, and the system can maximum detect four simultaneously arrival users. Essentially, we can design the parameters of a UDAS-based transmitter, to satisfy the requirement of a GFMA system.
Our proposed system can provide high spectrum efficiency, which can compare with a designed NOMA codebook.
For example, when $E_b/N_0$ = 7 dB and SE = 1.5, the BER of our proposed system is about $10^{-4}$, which is much better than that of the SCMA system.

Actually, the proposed UDAS-based MD-BICM transmitter can be extended to a general random access scenario, e.g., uncoordinated multiple-access. 
Moreover, there are many approaches to construct various UDAS sets. These contents are left as the future works.


\vfill
\end{document}